\documentclass[12pt,elsart,a4,amsmath]{elsart}
\usepackage{graphicx}
\usepackage{amsmath}
\usepackage{amssymb}
\begin{document}
\begin{frontmatter}
%\begin{titlepage}
\title{QED theory of transition probabilities 
and line profiles in highly charged ions}
\author[lab1,lab2]{L.N. Labzowsky},
\author[lab1]{ A. Prosorov},
\author[lab1]{ A. V. Shonin},
\author[lab3]{I. Bednyakov},
\author[lab3]{G. Plunien},
\and
\author[lab3]{G. Soff}
\address[lab1]{ St Petersburg University, Petrodvorets Uljanovskaya 1, 198904
St Petersburg, Russia }
\address[lab2]{Max-Planck-Institut f\"{u}r Physik Komplexer Systeme,
N\"{o}thnitzer Stra$\beta$e 38, D-01187 Dresden, Germany}
\address[lab3]{ Institut f\"{u}r Theoretische Physik, Technische
Universit\"{a}t Dresden, Mommsenstra$\beta$e 13, D-01062
Dresden, Germany }

\date{\today}

\begin{abstract}
A rigorous QED theory of the spectral line profiles is applied to
transition probabilities in few-electron highly charged ions. 
Interelectron interaction corrections are included as well as  
radiative corrections.  Parity nonconserving (PNC) amplitudes with
effective weak interactions between the electrons and nucleus are also
considered.  QED and interelectron interaction corrections to   the
PNC amplitudes are derived.

\end{abstract}
\end{frontmatter}

\clearpage
%\tableofcontents
%\clearpage
\section{Introduction}
\label{section1}

The spectra of the highly charged ions (HCI) are
currently under intensive experimental investigations~\cite{q1} -~\cite{qx}. Due to the strong field of the
nucleus acting on the electrons in HCI, the relevant theory should
be fully relativistic and based on the principles of QED. Up to now major experimental and theoretical efforts were
concentrated on the evaluation of the energy level shifts due to
radiative effects (Lamb shift) or to the interelectron
interaction, on the hyperfine splitting and on the bound electron
g-factor. The recent status of the problem was described in a
series of books and
reviews~\cite{qx+1},~\cite{qx+2},~\cite{qx+3},~\cite{qx+4}.

Transition probabilities were studied less intensively, though a
considerable amount of experimental~\cite{qx+5} -~\cite{qy} and
theoretical ~\cite{qy+1} -~\cite{qz} investigations also exists.
The full QED theory for the transition probabilities was
considered in ~\cite{qx+1},~\cite{qz+1} on the basis of the
S-matrix approach and in~\cite{qz+2} employing the two-time Green
function approach (see also~\cite{qx+4}).

In this paper we apply the most general QED line profile approach
to the derivation of expressions for transitions
probabilities in HCI with radiative, interelectron interactions and
weak interaction corrections.

The problem of the natural line profile in atomic physics was
considered first in terms of quantum mechanics by Weisskopf and
Wigner~\cite{qz+3}. In terms of modern QED it was first formulated
for one-electron atoms by Low~\cite{qz+4}. In~\cite{qz+4} the
appearance of the Lorentz profile in the resonance approximation
within the framework of QED was described and the nonresonant
corrections were estimated. Later the line profile QED theory was
modified also for few-electron ions ~\cite{qz+5} (see also
~\cite{qz+6}, ~\cite{qx+1}) and applied to the theory of
overlapping resonances in two-electron HCI
~\cite{qz+7},~\cite{qz+8}. Another application was provided to the
theory of nonresonant corrections in HCI and in neutral hydrogen
~\cite{qz+9},~\cite{qz+10},~\cite{qz+11}.

It was found in~\cite{qz+12} that the line profile approach
provides a convenient tool for calculating  energy corrections.
The most natural way to calculate the energy shift in QED is to
calculate the resonance shift in some scattering process. This way
clearly indicates the limits up to which the concept of the energy
of an exited states has a physical meaning -- that is the
resonance approximation. The exact theoretical value for the
energy of the exited state defined, for example, by the Green
function pole, can be compared directly with the measurable
quantities only in the resonance approximation when the line
profile is described by the two parameters: energy E and width
$\Gamma$. Beyond this approximation the evaluation of E and
$\Gamma$ should be replaced by the evaluation of the line profile
for the particular process. For applications of the line profile
approach to the calculation of energy level corrections see
~\cite{qz+13},~\cite{qz+14}.

In~\cite{qz+5},~\cite{qz+6} the line profile theory in QED was
developed on the basis of the evolution operator
$\hat{S}(0,\infty)$. Such a consideration, though most natural for
describing spontaneous emission and absorption processes, faces
serious difficulties due to unrenormalizability of the
corresponding S-matrix elements $<f|\hat{S}(0,\infty)|i>$. The
resonance approximation itself yields correct results 
~\cite{qz+5},~\cite{qz+6}, but the introduction of  radiative
corrections leads to the appearance of unrenormalizable
expressions.

Therefore in this paper we will use only renormalizable "full"
S-matrix elements   $<f|\hat{S}(\infty,-\infty)|i>$. 

During the last decades the parity nonconservation (PNC) processes in
HCI became a topic of permanent interest (up to now only from a
theoretical point of view)~\cite{qz+15} - ~\cite{qz+21},~\cite{qz+22}.

In neutral atoms, where successful experiments and accurate
theoretical calculations were performed, the main uncertainty is
introduced by electron correlation effects arising due to the
complicated atomic structure. In few-electron HCI electron
correlation does not play any significant role what greatly
simplifies the theoretical treatment. The major corrections to PNC
matrix elements in HCI are the radiative ones. This allows, in
principle, to test the Standard Model of electroweak interactions
beyond the "tree" level. It is especially important that the
experiments with HCI would provide these tests in the presence of
strong fields.

Partly the electroweak radiative corrections in HCI were
calculated in Ref.~\cite{qz+22}. In this work we present the full
treatment of the QED part of the electroweak corrections based on
the line profile approach. According to~\cite{qz+22} these
corrections are dominant in HCI and, unlike the so called
"oblique" corrections, are strongly field-dependent, i.e.
different in HCI and neutral atoms.

The paper is organized as follows. In section \ref{sec2} we begin with the
description of the process of  resonant photon scattering on
the "tree" level. 
In section \ref{sec3}  electron self-energy
insertions (SE) in the electron propagator within the resonance
approximation are considered. It is shown that they improve the
energy denominator of the photon emission (absorption) amplitude.
The same is done for the vacuum polarization (VP) insertions. In
 section \ref{sec4} the standard expressions for the Lorentz line
profile in the case of the photon emission (absorption) process is
derived; in section \ref{sec5} it is shown in which way higher-order
radiative corrections to the energy can be incorporated in the
Lorentz denominator. Section \ref{sec6a} is devoted to the derivation of 
radiative (QED) corrections to the initial state in the photon
emission amplitude.

The derivation of QED corrections to the final state requires
more refined considerations, that are contained in Section \ref{sec6}.
Here the process of  double resonance photon scattering is
introduced. In sections \ref{sec7}, \ref{sec8} the SE and VP insertions in the
different electron propagators in the Feynman diagram describing the
double photon scattering are considered. These insertions produce
 energy shifts to both resonant energy denominators of the
two-photon emission amplitude. In section \ref{sec9} the Lorentz profile
for the one-photon emission process in the case of an unstable finite
state is derived from the two-photon emission
amplitude. This derivation includes the integration over the
frequency of one of the photons. The resulting expressions can also be
used to include the Lamb shift of the final state
in the one-photon Lorentz denominator. The width of the final ground
state formally can be set equal to
zero; this approach can be considered as a kind of regularization
of the singular amplitudes. Section \ref{sec10} is devoted to the derivation
of QED corrections to the photon emission amplitude (final state).

In section \ref{sec11} the vertex correction to the photon emission
amplitude is evaluated; it is shown, that the ultraviolet and
infrared divergencies, contained in the vertex cancel with the
corresponding divergencies in the so called "derivative"
corrections.

 QED corrections to the amplitude admixed by the effective PNC
weak potential are investigated in Sections \ref{sec12}-\ref{sec14}. It is
supposed that the mixed states with opposite parity are separated
only by the Lamb shift and it is shown how the vanishing
denominators in the approximation of noninteracting electrons
acquire  a nonvanishing value. The cancellation of the
divergencies in the vertex and derivative graphs with the
effective PNC potential is also demonstrated.

In sections \ref{sec15}-\ref{sec18} the results obtained earlier are
generalized to few-electron ions. The first-order interelectron
interaction is taken into account explicitly. The introduction of
higher-order interelectron interaction corrections is discussed.
 First-order interelectron interaction corrections to the PNC
amplitude in the two-electron HCI are derived. Section \ref{sec19}
presents a short summary of all the results obtained in preceeding
sections.

\section{Resonant scattering of a photon on an atomic electron}
\label{sec2}
In this section we consider first the process of elastic photon
scattering on an atomic electron in lowest-order QED
perturbation theory. As usual for bound-electron QED we employ the
Furry picture in which the electrons are described by the solution of
the Dirac equation
\begin{equation}
\label{q0} (p\!\!\! /-m+\gamma_{0}eV)\Psi=0 \,.
\end{equation}
Here $p \!\! /\equiv\gamma_{\mu}p^{\mu},
p^{k}\equiv-i\nabla(k=1,2,3), p_{0}=E$, where $m$ and $e$ are the
electron mass and charge, respectively, $E$ is the energy,
$\gamma_{\mu}$ are the Dirac matrices and $V$ is the potential of
the nucleus (point-like or extended). The pseudoeuclidean metric (+ - -
-) in 4-space is used. We employ  relativistic units
$\hbar=c=1$ throughout the paper. The Feynman graph corresponding
to the process under consideration is presented in Fig.~1. We will
consider the elastic scattering and will assume that the initial
(final) state $A$ is the ground state. According to the standard
correspondence rules (see, for example,~\cite{qx+1}) the S-matrix
element for the graph Fig.~1 is:
\begin{equation}
\label{q1}   S^{(2)}=e^{2}\int d^4x_1d^4x_2
\bar{\Psi}_A(x_1)\gamma_{\mu_1}A_{\omega}^{\mu_1*}S(x_1,x_2)
\gamma_{\mu_2}A^{\mu_2}_{\omega'}(x_2)\Psi_A(x_2) \, ,
\end{equation}
where $\Psi_A(x)=\Psi_A(\vec{x}\,) e^{-iE_At}$ is the
electron wave function, $\bar{\Psi}_A$ is the Dirac conjugated
wave function, $A_{\omega}^{\mu_1*}(x)=
A_{\omega}^{\mu_1*}(\vec{x})\, e^{-i\omega t}$ is the wave
function (electromagnetic potential) for an absorbed photon with
 frequency $\omega$. The electron propagator for  bound
electrons is taken in the form:
\begin{equation}
\label{q2} 
  S(x_1,x_2)=\frac{1}{2\pi
i}\int\limits_{-\infty}^{\infty}e^{i\omega_1(t_1-t_2)}
\sum_n\frac{\bar{\Psi}_n(\vec{x}_1)\Psi_n(\vec{x}_2)}{E_n(1-i0)+\omega_1}d\omega_1 \, ,
\end{equation}
where the sum runs over the whole Dirac spectrum.
Integrating over time and frequency variables $t_1, t_2, \omega_1$
and using the relation between the S-matrix and the amplitude
$U$
\begin{equation}
\label{q3}  
S_{if}=-2\pi i\,\delta(E_i-E_f)U_{if}
\end{equation}
we will obtain for the scattering amplitude
\begin{equation}
\label{q4}
U^{(2)}_{\rm sc}=\sum_n
\frac{(U^{*}_{\omega})_{An_1}(U_{\omega'})_{n_1A}}{E_{n_1}-E_A-\omega}
\end{equation}
with the condition $\omega=\omega'$, which implies  energy
conservation. Here we abbreviate
\begin{equation}
\label{q5} 
U_{\omega}(\vec{x})\equiv
e\gamma_{\mu}A^{\mu}_{\omega}(\vec{x}) \, .
\end{equation}
In the resonance approximation the photon frequency $\omega$ is
close to the energy difference of two atomic levels: $\omega\approx
E_{A'}-E_A$. Accordingly, we have to retain only one term in the sum over $n$
in Eq. (\ref{q4})
\begin{equation}
\label{q6}
U^{(2)\rm{res}}_{\rm{sc}}=
\frac{(U^{*}_{\omega})_{AA'}(U_{\omega'})_{A'A}}{E_{A'}-E_A-\omega}\, .
\end{equation}
Eq. (\ref{q6}) shows that in the resonance approximation the
scattering amplitude factorizes into an emission and an absorption part.
It follows from Eq. (\ref{q5})
that the emission amplitude can be expressed as
\begin{equation}
\label{q7}
U_{\rm{em}}=\frac{(U^{*}_{\omega})_{AA'}}{E_{A'}-E_A-\omega}\, .
\end{equation}
For the absorption  we may write 
\begin{equation}
\label{q8}
U_{\rm{ab}}=\frac{(U_{\omega'})_{A'A}}{E_{A'}-E_A-\omega'} \,.
\end{equation}
This implies that the Lorentz profiles for  emission
and for  absorption will be identical as it should be.
Eq.~(\ref{q6}) also indicates the limit, up to which the
absorption and emission processes are independent - that is the
resonance approximation. Only within the resonance approximation
the spontaneous emission rate is independent from the process how the exited  state is prepared.

This approximation is quite adequate in atomic physics. Nonresonant corrections to the transition frequencies are still
too small to be observed in present modern 
experiments~\cite{qz+9} -~\cite{qz+11}. 

\section{Radiative insertions in the electron propagator}
\label{sec3}

In this section we consider  radiative insertions in the
electron propagator. The electron self-energy insertion
is depicted in Fig.~2. The S-matrix element corresponding to this diagram can be expressed as
\begin{multline}
\label{q9}
S^4 = e^4\int d^4x_1\ldots d^4x_4
\,
\bar{\Psi}_A(x_1)\gamma_{\mu_1}A^{\mu_1\ast}_{\omega}(x_1)S(x_1,x_2) \\
\times\gamma_{\mu_2}S(x_2,x_3)\gamma_{\mu_3}
D^{\mu_2\mu_3}(x_2,x_3)
S(x_3,x_4)\gamma_{\mu_4}A^{\mu_4}_{\omega'}(x_4)\Psi_A(x_4) \, ,
\end{multline}
where $D^{\mu_1\mu_2}(x_1,x_2)$ denotes the photon propagator
in  Feynman gauge
\begin{equation}
\label{q10}
 D^{\mu_1\mu_2}(x_1,x_2)=
\frac{1}{2\pi i}\frac{\delta^{\mu_1\mu_2}}{r_{12}}
\int\limits_{-\infty}^{\infty}d\omega \, 
e^{i\omega(t_1-t_2)+i|\omega|r_{12}}
\end{equation}
with $r_{12}=|\vec{x}_1-\vec{x}_2|$.
Integrating over  time and frequency variables and using again
Eq.~(\ref{q3}), we obtain the following  expression for the correction to the
scattering amplitude
\begin{equation}
\label{q11}
U^{(4)}_{\rm sc}=-\sum_{n_1,n_3}\frac{(U^*_{\omega})_{An_1}
(\hat{\Sigma}(E_A+\omega))_{n_1n_3}(U_{\omega'})_{n_3A}}
{(E_{n_1}-E_A-\omega)(E_{n_3}-E_A-\omega)} \, ,
\end{equation}
where $\hat{\Sigma}(E)$ is the electron self-energy
operator defined by its matrix elements
\begin{equation}
\label{q12} (\hat{\Sigma}(E))_{n_1n_3}=\frac{e^2}{2\pi
i}\sum_{n_2}\left(\frac{\gamma_{\mu_1}\gamma^{\mu_2}}{r_{12}}I_{E_{n_2}-E}(r_{12})
\right)_{n_1n_2n_2n_3} \, ,
\end{equation}
together with
\begin{equation}
\label{q13}
I_{E_{n_2}-E}(r_{12})=-\int\limits_{-\infty}^{\infty}\frac{d\omega
\,
e^{i|\omega|r_{12}}}{E_{n_2}-E+\omega} \,.
\end{equation}
In the resonance approximation  we have $n_1=n_3=A'$ and the
correction to the scattering amplitude takes the form
\begin{equation}
\label{q14}
U^{(4)\rm{res}}_{\rm{sc}}=-U^{(2)\rm{res}}_{\rm{sc}}\,
\frac{(\hat{\Sigma}(E_A+\omega))_{A'A'}} {(E_{A'}-E_A-\omega)} \,.
\end{equation}
Repeating the insertions in the resonance approximation
(the next term of this series is shown in Fig.~3) leads to a
geometric progression. Resummation of this progression
yields~\cite{qz+4}
\begin{equation}
\label{q15}
U^{\rm{res}}_{\rm{sc}}=\frac
{(U^{*}_{\omega})_{AA'}(U_{\omega'})_{A'A}}
{(\tilde{E}_{A'}-E_A-\omega)} \, ,
\end{equation}
where
\begin{equation}
\label{q16} \tilde{E}_{A'}=E_A+(\hat{\Sigma}(E_A+\omega))_{A'A'} \, .
\end{equation}

Accordingly, in the resonance approximation the emission amplitude is
 represented  by the expression
\begin{equation}
\label{q17}
U_{\rm{em}}=\frac{(U^{*}_{\omega})_{AA'}}{\tilde{E_{A'}}-E_A-\omega}\,.
\end{equation}
The operator $\hat{\Sigma}(E_A+\omega)$ may be expanded
into a Taylor series around the value $E_A+\omega=E_{A'}$:
\begin{equation}
\label{q18}
\hat{\Sigma}(E_A+\omega)=\hat{\Sigma}(E_{A'})+(E_{A'}-E_A-\omega)\Sigma'(E_{A'})+\ldots \, ,
\end{equation}
where
$\Sigma'(E_{A'})\equiv 
(\frac{\partial\hat{\Sigma}(E)}{\partial E})_{E=E_{A'}}$. 
The first two terms of the expansion (\ref{q18})
are ultraviolet divergent and have to be renormalized. The
renormalization of the first term to all orders in $\alpha Z$ is
well known. A commonly employed covariant procedure has been developed by
Mohr ~\cite{qz+23} (see further developments in ~\cite{qx+2}).
Within this approach the divergencies are canceled analytically.
Originally Mohr's method was first applicable numerically to high
and intermediate $Z$ values, but recently the high accurate
extension for $Z=1$ was also carried out~\cite{qz+24}. Another
noncovariant approach is based on the partial wave expansion of
$\hat{\Sigma}$ together with a numerical cancellation of divergencies for
each partial wave has been proposed in~\cite{qz+25},~\cite{qz+26}.
This approach is known as  partial wave renormalization (PWR). A
variant of the PWR where the divergencies are canceled partly
analytically and partly numerically utilizes the multicommutator expansion
method ~\cite{qz+27},~\cite{qz+28}.

Apart from self energy (SE) also the vacuum-polarization (VP) insertions in the electron propagator of
Fig.~1 should be considered to all orders in  resonance
approximation. The lowest order VP insertion is depicted in Fig.~4a. 
For the VP correction the Uehling approximation is frequently applied.
 In this approximation the bound-electron propagator in the
VP loop (Fig.~4a) is expanded in powers of the nuclear potential
( in powers of $\alpha Z$) and only the first term of the
expansion is retained (Fig.~4b). The Uehling approximation leads already to
 fairly good results even for high $Z$ values~\cite{qx+2}.
We will adopt this approximation  in our derivations for simplicity.
Accordingly, the VP insertions leads to the following modification of
Eq.~(\ref{q16}):
\begin{equation}
\label{q19}
\tilde{E}_{A'}=E_A+(\hat{\Sigma}(E_A+\omega))_{A'A'}+(V_U)_{A'A'} \, ,
\end{equation}
where $V_U$ is the Uehling potential~\cite{qz+29}.

\section{Line profile for the emission process}

\label{sec4}
In order to  obtain the line profile for the emission process we retain the
first term of the Taylor expansion (\ref{q18}) and consider the
energy denominator in Eq.~(\ref{q17}) with a shifted energy value
\begin{equation}
\label{q20}
\tilde{E}_{A'}=E_{A'}+(\hat{\Sigma}(E_{A'}))_{A'A'}+(V_U)_{A'A'} \, ,
\end{equation}
where
\begin{equation}
\label{q21}
(\hat{\Sigma}(E_{A'}))_{A'A'}=L^{\rm SE}_{A'}-\frac{i}{2}\Gamma_{A'}\, .
\end{equation}
$L^{\rm SE}_{A'}$ denotes the lowest-order electron self-energy contribution to   the
Lamb shift and $\Gamma_{A'}$ denotes the lowest order radiative width.
The other lowest-order correction to the Lamb shift is the vacuum
polarization 
\begin{equation}
\label{q22} (V_U)_{A'A'}=L^{\rm VP}_{A'} \, .
\end{equation}
Since the corresponding energy shift is real the vacuum polarization does not contribute to the
width. To this approximation the emission amplitude reads
\begin{equation}
\label{q23} U_{\rm{em}}=\frac{(U_{\omega}^{*})_{AA'}}
{E_{A'}+L_{A'}-E_A-\omega-\frac{i}{2}\Gamma_{A'}} \, ,
\end{equation}
where $L_{A'}=L_{A'}^{{\rm SE}}+L^{{\rm VP}}_{A'}$.
A method to incorporate the Lamb shift and the finite width of the state $A$ in
the energy denominator of Eq.~(\ref{q23}) will be discussed below in
section ~\ref{sec9}.

As the next steps of the calculation  we  have to take the square modulus of $U_{\rm{em}}$ and  to
integrate over the emission directions $\vec{\nu}$ of the photon
and to sum over the polarizations $\vec{e}$. Taking into account
the definition
\begin{equation}
\label{q24} \omega^{2}_{\rm{res}}\sum_{\vec{e}}\int
d\vec{\nu}\, |(U^*_{\omega})_{AA'}|^2=\Gamma_{AA'} \, ,
\end{equation}
where $\omega_{\rm{res}}$ is the resonant photon
frequency, $\Gamma_{AA'}$ is the partial width of the level $A'$,
associated with the transition $A'\rightarrow A$, we obtain an
expression for the transition probability $d W_{AA'}(\omega)$ for the
emission process
%\begin{multline}
\begin{eqnarray}
\label{q25}
dW_{AA'}(\omega)  &=&  \frac{1}{2\pi}
\sum_{\vec{e}}
\int
d\vec{\nu}\, |U_{\rm{em}}|^2 \omega^2 d\omega \nonumber \\ 
&=&  \frac{1}{2\pi}\frac{\Gamma_{AA'}d\omega}{(E_{A'}+L_{A'}-E_A-\omega)^2+
\frac{1}{4}\Gamma^2_{A'}}\, .
\end{eqnarray}
%\end{multline}
Eq.~(\ref{q25}) defines the Lorentz profile for the emission
spectral line. The values for the resonance frequency  in zeroth-order and in
first-order approximation read:
\begin{align}
\label{q26} \omega^{0}_{\rm{res}}&=E_{A'}-E_A ,
\\
\label{q27} \omega^{1}_{\rm{res}}&=E_{A'}+L_{A'}-E_A\, .
\end{align}
For further derivations it is more convenient to introduce a
complex  resonance frequency 
\begin{equation}
\label{q28}
\omega^{1}_{\rm{res}}=E_{A'}+(\hat{\Sigma}(E_{A'}))_{A'A'}-E_A\, .
\end{equation}

\section{Higher-order radiative corrections to the energy}
\label{sec5}

In this section we explain how  to incorporate higher-order
corrections to the energy levels in the denominator of
Eq.~(\ref{q25}). First, we have to take into account the next term of the
Taylor expansion (\ref{q18}). Considering this term as a
correction to the energy denominator, we have to insert  the resonance
frequency given by Eq.~(\ref{q28}). This leads to  the "reducible" part of the second-order 
loop-after-loop self-energy (SESE) correction~\cite{qx+1},~\cite{qz+30}.
\begin{equation}
\label{q29} \triangle
E_{A'}({\rm SESE,lal,red})=
(\hat{\Sigma}(E_{A'}))_{A'A'}(\hat{\Sigma}'(E_{A'}))_{A'A'}\, .
\end{equation}
The irreducible part of the SESE loop-after-loop
correction can be obtained according to  Fig.~3, where $n_2\neq A'$. The
corresponding contribution to the scattering amplitude is given by 
\begin{multline}
\label{q30}
U^{(6)}_{\rm se } = -\frac{(U^*_{\omega })_{AA'}}{E_{A'}-E_A-\omega}
\sum_{n_2\neq
A'}\frac{(\hat{\Sigma}(E_A+\omega))_{A'n_2}(\hat{\Sigma}(E_A+\omega))_{n_2A'}}
{E_{n_2}-E_A-\omega} \\
\times\frac{1}{E_{A'}-E_A-\omega}(U_{\omega})_{A'A} \, .
\end{multline}
In the sum over $n_2\neq A'$ we can set $E_A+\omega
=E_{A'}$. This sum can be viewed as a complicated higher-order
insertion in the electron propagator in Fig.~1. Repeating this
insertion subsequently in the resonance approximation, we
finally obtain the additional energy shift of the denominator of the
emission amplitude in Eq.~(\ref{q17}). This shift represents the
irreducible loop-after-loop SESE correction to the energy of the level $A'$:
\begin{equation}
\label{q31}
\triangle
E_{A'}({\rm SESE,lal,irr})=\sum_{n_2\neq A'}\frac
{({\hat \Sigma}(E_{A'}))_{A'n_2}({\hat \Sigma}(E_{A'}))_{n_2A'}}
{E_{n_2}-E_{A'}} \, .
\end{equation}
The corrections (\ref{q30}) and  (\ref{q31}) represent only two
of the second-order radiative corrections to the energy. The full
set of these corrections, i.e. the remaining second-order self-energy
(SESE) corrections, the second-order vacuum polarization (VPVP)
and the mixed SEVP corrections can be included in the denominator
of Eq.~(\ref{q17}) and hence in Eq.~(\ref{q25}) in the same way.
The expressions for these corrections can be found
in~\cite{qx+2} and ~\cite{qz+30}, respectively.

\section{ Radiative corrections to the emission amplitude (initial
state)}
\label{sec6a}

SE corrections to the amplitude (initial state), i.e. the SE
correction to the wave function $A'$ in the expression for the
amplitude follows according to the diagram in  Fig.~2 for intermediate states
 $n_1\neq A'$. 
 In this case the correction to $U_{\rm{sc}}$ takes the form
\begin{equation}
\label{q32}  
 U_{\rm{sc}}^{(4)}=-\sum_{n_1\neq A' }
 \frac{(U^*_{\omega})_{An_1}
 (\hat{\Sigma}(E_A+\omega))_{n_1A'}}
{E_{n_1}-E_A+\omega}\,
\frac{(U_{\omega})_{A'A}}{E_{A'}-E_A-\omega}  \,    .
\end{equation}
Here in the sum over $n_1$ we can set
$E_A+\omega=E_{A'}$. Comparing the expression ~(\ref{q32}) with
Eqs.~(\ref{q6}) and ~(\ref{q7}) we observe, that Eq.~(\ref{q32})
represents the corrections to the matrix element
$(U^*_{\omega})_{AA'}$ in the emission amplitude. Assuming, that
all the improvements  (corrections to the energy of the level  $A'$) in the denominator  are
already performed, we now can rewrite  formula (\ref{q17}) in the form
\begin{equation}
\label{q33}  
U_{\rm{em}}=\frac{(U^*_{\omega})_{A\tilde{A}'}}{\tilde{E_{A'}}-E_A-\omega}\, ,
\end{equation}
where
\begin{equation}
\label{q34} (U^*_{\omega})_{A\tilde{A}'}=\sum_{n_1\neq A'}
\frac{(U^*_{\omega})_{An_1}(\hat{\Sigma}(E_{A'}))_{n_1A'}}{E_{n_1}-E_{A'}}\, .
\end{equation}
The vacuum-polarization contribution can be introduced immediately into 
 Eq.~(\ref{q34}) 
\begin{equation}
\label{q35} (U^*_{\omega})_{A\tilde{A}'}=\sum_{n_1\neq
A'}\frac{(U^*_{\omega})_{An_1}[(\hat{\Sigma}(E_{A'}))_{n_1A'}+(V_U)_{n_1A'}]}{E_{n_1}-E_{A'}}
  \,   .
\end{equation}
In order to obtain the correction to the state $A$ in the matrix
element $(U^*_{\omega})_{AA'}$ as well as the correction to the 
 energy $E_A$ in the denominator of Eq.~(\ref{q33}), we have to
consider the diagrams  with  radiative insertions in the
upper external electron line. However, these graphs appear to be divergent. 
The singularity occurring can not be regularized in a direct manner.
To avoid this technical difficulty we may consider the double-photon-resonance
 scattering process of an electron in the state $A_0$ with
  resonance absorption into the state $A$ and finally into  the state
$A'$. If $A$ is the ground state, $A_0$ is some fictitious state
 that plays the role of a regulator. However the final expression for the
line profile for the emission transition probability
$A'\rightarrow A$ does not depend on $A_0$ and contains only a
dependence on the width $\Gamma_A$ of the state $A$. This width
can be set equal to zero at the end of the evaluations. Such a
regularization program will be carried out in the following sections.

\section{Double resonant photon scattering}
\label{sec6}

According to the idea discussed at the end of the previous
section, we now have to consider  the process depicted in the
Feynman diagram of Fig.~5. After the integration over time and
frequency variables the scattering amplitude corresponding to the
process depicted in Fig.~5 takes the form
\begin{multline}
\label{q36}
U^{(4)}_{\rm{sc}} = \sum_{n_1}\frac{(U^*_{\omega_0})_{A_0n_1}}
{E_{n_1}-E_{A_0}-\omega_0}
\sum_{n_2}\frac{(U^*_{\omega})_{n_1n_2}}{E_{n_2}-E_{A_0}-\omega_0-\omega}
\\
  \times
\sum_{n_3}\frac{(U_{\omega'})_{n_2n_3}(U_{\omega_0'})_{n_3A_0}}
{E_{n_3}-E_{A_0}-\omega_0-\omega+\omega'}\,  .
\end{multline}
The energy conservation during  this process is implemented by
the condition
\begin{equation}
\label{q37}   
\omega_{0}+\omega=\omega'+\omega_{0}'
\end{equation}
and the resonance frequencies are given by 
\begin{eqnarray}
\label{q38a}   
\omega_{0}=\omega_{0'}=E_A-E_{A_0} \, ,
\\
\label{q38b} 
\omega=\omega'=E_{A'}-E_{A} \, .
\end{eqnarray}
In the resonance case we have to set $n_1=A$, $n_2=A'$,
$n_3=A$ in Eq.~(\ref{q36}), which yields
\begin{equation}
\label{q39}
   U^{(4)\rm{res}}_{\rm{sc}}=\frac
{(U^*_{\omega_0})_{A_0A}
(U^*_{\omega})_{AA'}(U_{\omega'})_{A'A}(U_{\omega_0'})_{AA_0}}
{(E_A-E_{A_0}-\omega_0)(E_{A'}-E_{A_0}-\omega_0-\omega)(E_A-E_{A_0}-\omega_0')}
\, .
\end{equation}
In order to describe the line 
profile for  double-photon emission we have to
 consider 
\begin{equation}
\label{q40} 
  U_{\rm em}=\frac
{(U^*_{\omega_0})_{A_0A}(U^*_{\omega})_{AA'}}
{(E_A-E_{A_0}-\omega_0)(E_{A'}-E_{A_0}-\omega_0-\omega)} \, ,
\end{equation}
i.e.  the amplitude for double photon emission in
 resonance approximation. An analogous expression can be
derived for the amplitude of double photon absorption
\begin{equation}
\label{q41}  
 U_{ab}=\frac          
{(U_{\omega'})_{A'A}(U_{\omega_0'})_{AA_0}}
{(E_{A'}-E_{A_0}-\omega_0'-\omega')(E_A-E_{A_0}-\omega_0')}\, .
\end{equation}

\section{Radiative insertions in the central electron propagator}
\label{sec7}

We start with the graph depicted in Fig.~6. Having performed the integrations over time and
frequency  we find
\begin{multline} 
\label{q42}
U^{(6)}_{\rm{sc}} = \sum_{n_1} \frac{(U^*_{\omega_0})_{A_0n_1}}
{E_{n_1}-E_{A_0}-\omega_0}\sum_{n_2n_3}\frac
{(U^*_{\omega})_{n_1n_2}(\hat{\Sigma}(E_{A_0}+\omega_0+\omega))_{n_2n_3}}
{(E_{n_2}-E_{A_0}-\omega_0-\omega)(E_{n_3}-E_{A_0}-\omega_0-\omega)}
\\ \times \sum_{n_4}\frac
{(U_{\omega'})_{n_3n_4}(U_{\omega_0'})_{n_4A_0}}
{E_{n_4}-E_{A_0}-\omega_0-\omega+\omega'}\, .  
\end{multline} 
In the resonant case $(n_1=A, n_2=n_3=E_{A'},n_4=A)$ 
the equations above yields 
%\begin{multline} 
\begin{eqnarray}
\label{q43}
U^{(6)\rm{res}}_{\rm{sc}} &=& \frac
{(U^*_{\omega_0})_{A_0A}(U^*_{\omega})_{AA'}}
{(E_{A}-E_{A_0}-\omega_0)(E_{A'}-E_{A_0}-\omega_0-\omega)}\nonumber\\
&& \times \frac{(\hat{\Sigma}(E_{A_0}+\omega_0+\omega))_{AA'}}
{E_{A'}-E_{A_0}-\omega_0-\omega}\,
\frac{(U_{\omega'})_{A'A}(U_{\omega_0'})_{AA_0}}
{E_{A}-E_{A_0}-\omega_0'}\,  ,  
%\end{multline} 
\end{eqnarray}
where   the
condition (\ref{q37}) has been employed in the last denominator.  The resummation of the
SE and VP insertions to all orders in perturbation theory leads to
the following expression for the emission amplitude
\begin{equation}
\label{q44}   U_{\rm{em}}=\frac
{(U^*_{\omega_0})_{A_0A}(U^*_{\omega})_{AA'}}
{(E_{A}-E_{A_0}-\omega_0)(\tilde{E}_{A'}(\omega+\omega_0)-E_{A_0}-\omega_0-\omega)} \, ,
\end{equation}
where
\begin{equation}
\label{q45}  
\tilde{E}_{A'}(\omega+\omega_0)=E_{A'}+(\hat{\Sigma}(E_{A_0}+\omega_0+\omega))_{A'A'}+(V_U)_{A'A'}
\, .
\end{equation}
This result is similar to  Eq.~(\ref{q19}) which  has been derived for
the one-photon resonance case. We can also repeat a similar derivation as  
for the result
Eq.~(\ref{q33}). For this purpose we should take $n_1=A, n_2\neq A',
n_3=A', n_4=A$ in Fig.~6. Accordingly, we find  for the emission amplitude
\begin{equation}
\label{q46}   U_{\rm{em}}=\frac
{(U^*_{\omega_0})_{A_0A}(U^*_{\omega})_{A\tilde{A}'}}
{(E_{A}-E_{A_0}-\omega_0)(\tilde{E}_{A'}(\omega+\omega_0)-E_{A_0}-\omega_0-\omega)} \, ,
\end{equation}
where $(U^*_{\omega})_{A\tilde{A}'}$ is defined by
Eq.~(\ref{q35}). Eq.~(\ref{q46}), as well as Eq.~(\ref{q33}) 
describes the radiative correction to the emission amplitude that
improves the wave function of the initial state $A'$.

\section{Radiative insertions in the upper electron propagator}
\label{sec8}

In this section we turn to the radiative insertions into the upper electron
propagator of Fig.~5. To give an example the lowest-order SE insertion 
is depicted in Fig.~7. Performing again integrations over  time and frequency in the
corresponding S-matrix element yields
%\begin{multline}
\begin{eqnarray}
\label{q47}
U^{(6)}_{\rm{sc}} &=& \sum_{n_1n_2}\frac
{(U^*_{\omega_0})_{A_0n_1}(\hat{\Sigma}(E_{A_0}+\omega))_{n_1n_2}}
{(E_{n_1}-E_{A_0}-\omega_0)(E_{n_2}-E_{A_0}-\omega_0)}\nonumber\\
&& \times \sum_{n_3}\frac {(U^*_{\omega})_{n_2n_3}}
{(E_{n_3}-E_{A_0}-\omega_0-\omega)} \sum_{n_4}\frac
{(U_{\omega'})_{n_3n_4}(U_{\omega_0'})_{n_4A_0}}
{(E_{n_4}-E_{A_0}-\omega_0-\omega+\omega')} \,  .
\end{eqnarray}
%\end{multline}
The resonant case is determinated by the conditions: $n_1=A, n_2=A, n_3=A',n_4=A$,  i.e.
%\begin{multline}
\begin{eqnarray}
\label{q48}
U^{(6)\rm{res}}_{\rm{sc}} &=& 
\frac{(U^*_{\omega_0})_{A_0A}}{E_A-E_{A_0}-\omega_0}\,
\frac{(\hat{\Sigma}(E_{A_0}+\omega_0))_{AA}}{E_A-E_{A_0}-\omega_0}\nonumber\\
\times
\frac{(U^*_{\omega})_{AA'}}{E_{A'}-E_{A_0}-\omega_0-\omega}\,
\frac{(U_{\omega'})_{A'A}(U_{\omega_0'})_{AA_0}}{E_{A}-E_{A_0}-\omega_0'}
\, .
\end{eqnarray}
%\end{multline}
We can assume, that all the resonant radiative
insertions into the central electron propagator in Fig.~5 are
already taken into account.  Repeating the radiative insertions in
the upper electron line in resonance approximation and summing
up the resulting geometrical progression finally yields
\begin{equation}
\label{q49}  
 U_{\rm{em}}=\frac
{(U^*_{\omega_0})_{A_0A}(U^*_{\omega})_{AA'}}
{(\tilde{E}_A(\omega_0)-E_{A_0}-\omega)(\tilde{E}_{A'}(\omega_0+\omega)-E_{A_0}-\omega_0-\omega)} 
\end{equation}
together with
\begin{equation}
\label{q50}  
\tilde{E}_A(\omega_0)=E_A+(\hat{\Sigma}(E_{A_0}+\omega_0))_{AA}+(V_U)_{AA}\, .
\end{equation}
Eq.~(\ref{q49}) represents the expression for  the double-photon
 amplitude where both energies $E_A$ and $E_{A'}$ are
improved due to  radiative corrections.

\section{Line profile for the photon emission with energy
corrections to the final state}
\label{sec9}

For this  purpose it is sufficient to keep only the leading terms in
the Taylor expansion of the denominators in Eq.~(\ref{q49}):
\begin{equation}
\label{q51a}  
\tilde{E}_A(\omega_0)=E_A+(\hat{\Sigma_A}(E_A))_{AA}+(V_U)_{AA}=E_A+L_A-\frac{i}{2}\Gamma_A
\, ,
\end{equation}
\begin{equation}
\label{q51b}  
\tilde{E}_{A'}(\omega_0+\omega)=E_{A'}+(\hat{\Sigma_{A'}}(E_{A'}))_{A'A'}+(V_U)_{A'A'}
=E_{A'}+L_{A'}-\frac{i}{2}\Gamma_{A'}
\, .
\end{equation}
Integrating over both photon directions and summing over
the polarizations (see section \ref{sec4}) an expression for the
 transition probability~\cite{qz+31} of double-photon emission is obtained
%\begin{multline}
\begin{eqnarray}
\label{q52}   
dW_{A' \rightarrow A \rightarrow A_0}
 &=& \frac{1}{(2\pi)^2}
\frac{\Gamma_{A_0A}}{|E_A+L_A-E_{A_0}-\omega_0-\frac{i}{2}\Gamma_A|^2}
\nonumber\\ 
&& \times 
\frac{\Gamma_{AA'}\,d\omega d\omega_0}
{|E_{A'}+L_{A'}-E_{A_0}-\omega_0-\omega-\frac{i}{2}\Gamma_{A'}|^2} \, ,  
%\end{multline}
\end{eqnarray}
where $\Gamma_{A_0A}, \Gamma_{AA'}$ denote the partial
widths  as defined in section \ref{sec4}. In a next step we integrate over $\omega_0$
in the complex plane. The integration can be extended along the real axis
 $-\infty <\omega_0<\infty$ since only the residues at the poles are
contributing. We choose the contour in the upper half-plane. 
The poles are located at 
\begin{align}
\label{q53a}  
\omega^{(1)}_0&=\omega_{AA_0}+L_A+\frac{i}{2}\Gamma_A \, ,
  \\ 
\label{q53b} 
\omega^{(2)}_0&=\omega_{A'A_0}+L_{A'}-\omega+\frac{i}{2}\Gamma_{A'}
\end{align}
where $\omega_{AA_0}=E_A-E_{A_0}$ and 
$\omega_{A'A_0}=E_{A'}-E_{A_0}$, respectively.
As the result of the integration a new expression for transition probability $dW_{AA'}$ is derived, where the
Lamb shift and the width of the state $A$ are taken into account
%\begin{multline}
\begin{eqnarray}
\label{q54}  
dW_{AA'} &=& {\frac{1}{2\pi}}
\Gamma_{A_0A}
\Gamma_{AA'} \nonumber\\
&& \times \left\{
\frac{1}{\Gamma_A}
\frac{1}{|\omega_{A'A_0}+L_{A'}-\omega_{AA_0}-L_A-\frac{i}{2}(\Gamma_{A'}+\Gamma_{A})-\omega|^2}
\right. 
\nonumber\\ 
&& \left.+ 
\frac{1}{\Gamma_{A'}}
\frac{1}{|\omega_{AA_0}+L_{A}-\omega_{A'A_0}-L_{A'}-\frac{i}{2}(\Gamma_{A'}+\Gamma_{A})+\omega|^2}
\right\}d\omega  
\, .
%\end{multline}
\end{eqnarray}
After simple algebraic transformations 
Eq.~(\ref{q54}) can be cast into the form
\begin{equation}
\label{q55} 
  dW_{AA'}=\frac{1}{2\pi}
\frac{\Gamma_{A_0A}
\Gamma_{AA'}
}{\Gamma_{A}\Gamma_{A'}}
\frac{(\Gamma_A+\Gamma_{A'})\,d\omega}{(\tilde{\omega}_{A'A}-\omega)^2+\frac{1}{4}(\Gamma_{A'}+\Gamma_A)^2}
\end{equation}
where $\tilde{\omega}_{A'A}=E_{A'}+L_{A'}-E_A-L_A$. The latter 
 reveals how the Lamb shift $L_A$ of the final state $A$ enters the Lorentz
denominator in the expression for the emission probability.

For simplicity, we may assume that $\Gamma_{A_0A}=\Gamma_A$ and
$\Gamma_{AA'}=\Gamma_{A'}$. This implies that both states $A'$ and
$A$ have only one decay channel: $A'\rightarrow A$ and
$A\rightarrow A_0$. Finally, we arrive at 
\begin{equation}
\label{q56}   dW_{AA'}=\frac{1}{2\pi}
\frac{(\Gamma_A+\Gamma_{A'})\, d\omega}
{(\tilde{\omega}_{A'A}-\omega)^2+\frac{1}{4}(\Gamma_{A'}+\Gamma_A)^2}\,.
\end{equation}
This formula does not contain any dependence on the
state $A_0$. This state enters only indirectly through the definition
of $\Gamma_A$. If $A$ denotes the ground state, we can set
$\Gamma_A=0$ in Eq.~(\ref{q56}) which yields
\begin{equation}
\label{q57}   dW_{AA'}=\frac{1}{2\pi}
\frac{\Gamma_{A'} d\omega}
{(\tilde{\omega}_{A'A}-\omega)^2+\frac{1}{4}\Gamma_{A'}^2}\, .
\end{equation}
The expression above deviates  from Eq.~(\ref{q25}) only by the
presence of the Lamb shift $L_A$ in the denominator.  The
introduction of the state $A_0$ acts as a  regularization and
$\Gamma_A$ plays the role of the regularization parameter. In case
of the ground state $A$  this parameter is removed by setting it equal to zero at the
end of the derivation.

\section{Radiative corrections to the emission amplitude (final
state)}
\label{sec10}
Now we are in the position to determine the final-state radiative correction
to the emission amplitude, i.e. to correct the final-state wave
function. For this purpose we return to Fig.~7 and consider the
case: $n_1=A, n_2\neq A, n_3=A', n_4=A$. We assume that the
resummation of all  radiative insertions in resonance
approximation in Fig.~7 is already performed. Instead of
Eq.~(\ref{q49}) we can write
\begin{equation}
\label{q58}   U_{\rm{em}}=\frac
{(U^*_{\omega_0})_{A_0A}(U^*_{\omega})_{\tilde{A}A'}}
{(\tilde{E}_A(\omega_0)-E_{A_0}-\omega_0)(\tilde{E}_{A'}(\omega_0+\omega)-E_{A_0}-\omega_0-\omega)}\, ,
\end{equation}
where
\begin{equation}
\label{q59}  
(U^*_{\omega})_{\tilde{A}A'}=\sum_{n_2\neq A}\frac
{\left[(\hat{\Sigma}(E_A))_{An_2}+(V_U)_{An_2}\right](U^*_{\omega})_{n_2A'}}
{E_{n_2}-E_A} \, .
\end{equation}
Comparing Eq.~(\ref{q59}) with Eqs.~(\ref{q33}) and  (\ref{q35}) one
can see  that it represents the corrections to the
final-state wave function in the expression for the emission
amplitude.

Collecting now all the various corrections to the amplitudes and to energy
denominators given by Eqs.~(\ref{q35}), (\ref{q59}), (\ref{q19})
and (\ref{q50}), respectively, we can  finally write the generic equation
(\ref{q57}) into the form (for the ground state $A$):
\begin{equation}
\label{q60}   dW_{AA'}=\frac{1}{2\pi}\frac
{\Gamma_{\tilde{A}\tilde{A}'} d\omega}
{(\tilde{\omega}_{A'A}-\omega)^2+\frac{1}{4}\Gamma^2_{A'}}\,  .
\end{equation}
 For convenience we reintroduced  the notation
$\Gamma_{A'}=\Gamma_{AA'}$ and denoted by
$\Gamma_{\tilde{A}\tilde{A}'}$ the expression for the width
$\Gamma_{A'}$ (i.e. for the transition rate $A'\rightarrow A$)
where both wave functions of the initial $(A')$ and final $(A)$
states are corrected according to the formulas (\ref{q35}) and
(\ref{q59}). These corrections to the width are additive, so that
\begin{equation}
\label{q61}  
\Gamma_{\tilde{A}\tilde{A}'}=\Gamma_{AA'}+\Gamma_{\tilde{A}A'}+\Gamma_{A\tilde{A}'}
\end{equation}
where $\Gamma_{\tilde{A}A'}$ and $\Gamma_{A\tilde{A}'}$
denote the corrections to $\Gamma_{AA'}$ which contain the improved
amplitudes (\ref{q59}) and (\ref{q35}).

In principle, the width $\Gamma_{A'}$ in the denominator of
Eq.~(\ref{q60}) also should be replaced by
$\Gamma_{\tilde{A}\tilde{A}'}$. This means, that a rigorous
evaluation of the radiative corrections to the transition
probability should include the second-order radiative corrections
in the denominator of Eq.~(\ref{q60}). The imaginary parts of
these second-order corrections would determine exactly the value of
$\Gamma_{\tilde{A}\tilde{A}'}$~\cite{qx+1}.

\section{Vertex correction}
\label{sec11}

A third type of  QED corrections to the emission amplitude, which we have
to consider are  vertex corrections. For the evaluation of
these corrections we can return at first to the one-photon resonance
picture (Eq.~(\ref{q17})). The SE vertex correction to the
emission amplitude corresponds to the Feynman graph shown in Fig
8. In the resonance approximation this correction reads
\begin{equation}
\label{q62}  
U^V_{\rm{em}}=\frac{(\Lambda^{(1)}_{\mu}A^{\mu
*}_{\omega})_{AA'}}{\tilde{E}_{A'}-E_A-\omega} \, ,
\end{equation}
where $\Lambda^{(1)}_{\mu}$ is the SE vertex that is
defined directly by the upper part of the diagram in Fig.~8. The
renormalization of $\Lambda^{(1)}_{\mu}$ may be performed in
momentum space~\cite{qz+30}
\begin{equation}
\label{q63}   \Lambda^{(1)\rm{ren}}_{\mu}(\not p',
\not p)=\Lambda^{(1)}_{\mu}(\not p', \not
p)-\gamma_{\mu}\Lambda^{(1)} \, ,
\end{equation}
where
\begin{equation}
\label{q65}  
 \gamma_{\mu}\Lambda^{(1)}=-4\pi
ie^2\int\frac{d^4k}{(2\pi)^4}\gamma_{\nu}\frac{\not
k}{k^2}\gamma_{\mu}\frac{\not k}{k^2}\gamma^{\nu}\frac{1}{k^2} \, .
\end{equation}
In Feynman gauge the counterterm $\gamma_{\mu}\Lambda^{(1)}$ is
ultraviolet and infrared divergent. This divergency cancels, if 
we take into account another contribution to the SE vertex that
follows from the graph Fig.~2  and which has not yet been considered.
Without changing the results of Section~\ref{sec3} we can take the first
term of the geometric progression with SE corrections and multiply
the expression without radiative corrections (i.e., the right-hand
side of Eq.~(\ref{q17})) by this term:
\begin{equation}
\label{q66}  
U^o_{\rm{em}}+U^d_{\rm{em}}=-\frac{(\gamma_{\mu}A^{\omega*}_{\mu})_{AA'}}{\tilde{E}_{A'}-E_A-\omega}
\,\frac{(\hat{\Sigma}(E_A+\omega))_{A'A'}}{E_{A'}-E_A-\omega} \, .
\end{equation}

Here we employ  the notation $U^o_{\rm{em}}$ for the
emission amplitude Eq.~(\ref{q17}) without radiative corrections
and $U^d_{\rm{em}}$ stands for "derivative" correction that will
follow from Eq.~(\ref{q66}).

Using the Taylor expansion (\ref{q18}) for $\Sigma(E_A+\omega)$
and substituting the
resonant value (\ref{q28}) for the frequency 
$\omega=\omega^{(1)}_{\rm{res}}=E_{A'}+\Sigma_{A'}-E_A$ into the energy denominator in Eq.~(\ref{q66}), we obtain
\begin{equation}
\label{q67}  
U^o_{\rm em }+U^d_{\rm{em}}=-\frac{(\gamma_{\mu}A^{\omega*}_{\mu})_{AA'}}{\tilde{E}_{A'}-E_A-\omega}
\left[1+(\hat{\Sigma}'(E_{A'}))_{A'A'}\right]\, ,
\end{equation}
respectively 
%\begin{multline}
\begin{eqnarray}
\label{q68}
U^o_{\rm em }+U^d_{\rm em}+U^V_{\rm em} &=&
-\frac{(U^*_{\omega})_{AA'}}{\tilde{E}_{A'}-E_A-\omega} \nonumber\\
&& +\frac{1}{\tilde{E}_{A'}-E_A-\omega}
\left[(\Lambda^{(1)}_{\mu}A^{\mu*}_{\omega})_{AA'}+
\Sigma'_{A'}(\gamma_{\mu}A^{\mu*}_{\omega})_{AA'}\right] \, .
%\end{multline}
\end{eqnarray}
The renormalized expression for $\hat{\Sigma}'$ is given by \cite{qz+30}
\begin{equation}
\label{q69} {\hat{\Sigma'}}^{\rm{ren}}
(\not p)={\hat{\Sigma'}}(\not p)-\Sigma' \, ,
 \end{equation}
where
\begin{equation}
\label{q70} 
\left.\gamma_{\mu}
\Sigma'
\equiv
\frac{\partial}{\partial p_{\mu}}
\hat{\Sigma}(\not p)
\right|_{\not p=m}
\end{equation}
and $\hat{\Sigma}(\not{ p})$ denotes the lowest-order electron
self-energy operator in  momentum space. The counterterm $\Sigma'$
contains ultraviolet and infrared divergencies. Due to the Ward
identity
\begin{equation}
\label{q71}   
\Lambda^{(1)}=-\Sigma'
\end{equation}
all the divergencies in Eq.~(\ref{q68}) cancel. The total
renormalized expression Eq.~(\ref{q68}) seems to be asymmetric with
respect to $A$ and $A'$. A symmetric result can be obtained from
the two-photon resonance scattering.

Consider now Eq.~(\ref{q49}). Repeating similar
derivations as have been performed above for Eq.~(\ref{q17}), we obtain
\begin{multline}
\label{q72}  
U^o_{\rm{em}}+U^d_{\rm{em}}=
\frac{1}{2}
\frac{(U^*_{\omega_0})_{A_0A}(U^*_{\omega})_{AA'}}
{(\tilde{E}_{A}(\omega_0)-E_{A_0}-\omega)(\tilde{E}_{A'}
(\omega_0+\omega)-E_{A_0}-\omega_0-\omega)}
 \\ 
\times\left\{\left(-\frac{\Sigma_{A'}}{E_{A'}-E_{A_0}-\omega^{(0)}_{0\rm res}
-\omega^{(1)}_{\rm res }+{\Sigma'}^{(\rm ren)}_{A'}}
\right)
\right.
 \\
+
\left.
\left(-\frac{\Sigma_{A'}}{E_{A}-E_{A_0}-\omega^{(1)}_{0\rm res}+{\Sigma'}^{(\rm ren)}_{A}}\right)\right\}
\,  ,
\end{multline}
together with the lowest-order resonance energies
%\begin{align}
\begin{eqnarray}
\label{q73a}   
\omega^{(0)}_{0\rm res } &=& E_{A}-E_{A_0} \,  ,
\\
\label{q73b}  
\omega^{(1)}_{0 \rm res }&=& E_{A}-E_{A_0}+\Sigma_A \, ,
\\
\label{q73c}  
\omega^{(1)}_{\rm{res}} &=& E_{A'}-E_{A}+\Sigma_{A'} \, .
%\end{align}
\end{eqnarray}
We omit here VP contributions in the denominators of
Eq.~(\ref{q72}). The values
$\Sigma_A\equiv(\hat{\Sigma}(E_A))_{AA}$ and
$\Sigma_{A'}\equiv(\hat{\Sigma}(E_{A'}))_{A'A'}$ are considered as being
renormalized $\Sigma_{A}=\Sigma^{\rm (ren) }_{A}$,
$\Sigma_{A'}=\Sigma^{\rm{(ren)}}_{A'}$.  Substitution of
Eqs.~(\ref{q73a}), (\ref{q73b}), (\ref{q73c}) into Eq.~(\ref{q72})
yields a symmetrized expression for the sum:
\begin{multline}
\label{q74}
U^o_{\rm em }+U^d_{\rm em } = 
\frac{(U^*_{\omega_0})_{A_0A}(U^*_{\omega})_{AA'}}{(\tilde{E}_{A}(\omega_0)-E_{A_0}-\omega)(\tilde{E}_{A'}
(\omega_0+\omega)-E_{A_0}-\omega_0-\omega)}
  \\
  \times \left(1+\frac{1}{2}\left[\Sigma^{'\rm{(ren)}}_{A'}+
{\Sigma'}^{\rm{(ren)}}_{A}\right]\right)
\, .
\end{multline}
The cancellation of the divergencies with those in $U^V_{\rm{em}}$ still
takes place.

Now the expression (\ref{q61}) requires to correct the width  by replacing
$\Gamma_{\tilde{A}\tilde{A}'}$ by
$\tilde{\Gamma}_{\tilde{A}\tilde{A}'}$ where
\begin{equation}
\label{q75}   \tilde{\Gamma}_{\tilde{A}\tilde{A}'}=
\tilde{\Gamma}_{AA'}+{\Gamma}_{\tilde{A}A'}+
{\Gamma}_{A\tilde{A}'} \, .
\end{equation}
The width $\tilde{\Gamma}_{AA'}$ results from Eq.~(\ref{q25}) by  replacing the amplitudes $U_{\rm{em}}$ by the modified
expressions Eq.~(\ref{q68}) and  (\ref{q74}), respectively.

What further remains is to include the contribution of the VP vertex as  depicted
in Fig.~9. In  Uehling approximation the renormalization of the
corresponding correction is straightforward. One has to replace
the expression for the electron loop in Fig.~9 b), i.e. the photon
self energy $\Pi(x_1x_2)$ by the known renormalized expression
$\Pi_R$~\cite{qz+31}.

\section{Weak interaction mixed amplitude}
\label{sec12}

To begin the investigation of additional corrections to the amplitude due to  
weak interaction we return to the generic one-photon scattering Feynman
diagram Fig. 1 by adding at  first the effective weak-interaction potential in
the electron propagator (see Fig. 10). We can specify the
corresponding scattering amplitude as
\begin{equation}
\label{q76} 
  U_{\rm{sc}}=\sum_{n_1}\frac
{(U^*_{\omega})_{An_1}}{E_{n_1}-E_A-\omega} \sum_{n_2}\frac
{(\gamma_0V_{\rm{PNC}})_{n_1n_2}(U_{\omega'})_{n_2A}}
{E_{n_2}-E_A-\omega} \, ,
\end{equation}
together with the effective  parity-nonconserving (PNC) potential
\begin{equation}
\label{q77}  
V_{\rm{PNC}}=-\frac{G_F}{2\sqrt{2}}Q_{w}\rho_N(r)\gamma_5 \, .
\end{equation}
$Q_w$ denotes the "weak charge" of the nucleus
\begin{equation}
\label{q78}   Q_w=Z(1-4\sin^2\Theta_w)-N \, ,
\end{equation}
which relates the numbers of protons $Z$ and neutrons $N$ in the
nucleus and  the Weinberg angle  $\Theta_w$ - the free parameter
of the electroweak theory. The latter is usually determined via a comparison between
theoretical predictions and data from  various experiments both in
high-energy and atomic physics. The recent adopted value for the Weinberg angle
 is $\Theta_w=0.2312$~\cite{qz+32}.

Within the resonance approximation we set $n_2=A'$ in Eq. (\ref{q76}). Since the operator
$V_{\rm{PNC}}$ has no diagonal matrix elements the term $n_1=A'$
is absent. However, we can choose  a level $n_1=A''$ which gives a dominant
contribution to the sum over $n_1$. A standard example in atomic
physics is: $A=1s, A'=2s, A''=2p$. Accordingly in the sum over $n_1$ the
term yielding the  small denominator $E_{A''}-E_{A'}$ (in the resonance
approximation)  dominates. The emission amplitude reads
\begin{equation}
\label{q79a}  
U_{\rm{em}}=\frac{(U^*_{\omega})_{AA''}}{E_{A''}-E_A-\omega}
\,\frac{(\gamma_0V_{\rm PNC})_{A''A'}}{E_{A'}-E_A-\omega} \, .
\end{equation}
Introducing the resonance frequency  in the first denominator we
can further write
\begin{equation}
\label{q79b}  
U_{\rm{em}}=\frac{(U^*_{\omega})_{AA''}}{E_{A''}-E_{A'}}
\,\frac{(\gamma_0V_{\rm{PNC}})_{A''A'}}{E_{A'}-E_A-\omega}\, .
\end{equation}
In particular for states $A'=2s, A''=2p$ in the approximation of noninteracting
electrons $E_{A''}=E_{A'}$. This implies that in  the first denominator in
Eq. (\ref{q79b}) one has to account for the Lamb shift $\Delta E_L$.
In the next section we will show how it arises and how to  use
 the form (\ref{q79a}) for $U_{\rm{em}}$ for this purpose.

\section{Radiative corrections to the weak interaction mixed
amplitude}
\label{sec13}

The first step consist in the evaluation of  the SE insertions in the upper
electron propagator in the diagram of Fig.~10 within the resonance approximation (see Fig.~11 for
the first term of this sequence).

The scattering amplitude, corresponding to the diagram of Fig.~11 is given by
%\begin{multline}
\begin{eqnarray}
\label{q80}
U_{\rm{sc}} = \sum_{n_1}\frac{(U^*_{\omega})_{An_1}}{E_{n_1}-E_A-\omega}
 \sum_{n_2n_3}\frac{(\hat{\Sigma}(E_A+\omega))_{n_1n_2}
(\gamma_0V_{\rm{PNC}})_{n_2n_3}(U_{\omega'})_{n_3A}}
{(E_{n_2}-E_A-\omega)(E_{n_3}-E_A-\omega)}\, .
\end{eqnarray}
%\end{multline}
The resonant state is  $n_3=E_{A'}$. The states yielding dominant contributions are
$n_1=n_2=E_{A''}$. Keeping only these terms in the summations  we obtain
\begin{equation}
\label{q81}  
U_{\rm{em}}=\frac{(U^*_{\omega})_{AA''}}{E_{A''}-E_A-\omega}
\,\frac{(\hat{\Sigma}(E_A+\omega))_{A''A''}}{(E_{A''}-E_A-\omega)}\,
\frac{(\gamma_0V_{\rm{PNC}})_{A''A'}}{E_{A'}-E_A-\omega}
\, .
\end{equation}
In  resonance approximation we can set $E_A+\omega=E_{A'}$ in
the matrix element $(\hat{\Sigma}(E_A+\omega))_{A''A''}$ and,
furthermore,  $E_{A'}=E_{A''}$ due to
the almost degeneracy of the levels $A'$ and $A''$. Then the
resummation of the SE insertions yields
\begin{equation}
\label{q82}
U_{\rm{em}}=\frac{(U^*_{\omega})_{AA''}}{\tilde{E}_{A''}-E_A-\omega}
\,\frac{(\gamma_0V_{\rm{PNC}})_{A''A'}}{E_{A'}-E_A-\omega} \, .
\end{equation}
If $n_1=A''$ but $n_2\neq A''$, we obtain the correction to the
PNC matrix element $(V_{\rm{PNC}})_{A''A'}$ (assuming that 
all insertions in the resonant approximation are already performed)
\begin{align}
\label{q83}  
U_{\rm{em}}&=\frac{(U^*_{\omega})_{AA''}}{\tilde{E}_{A''}-E_A-\omega}
\,\frac{(\gamma_0V_{\rm{PNC}})_{\tilde{A}''A'}}{E_{A'}-E_A-\omega} \, ,
\\
\label{q84}  
(\gamma_0V_{\rm{PNC}})_{\tilde{A}''A'}&=\sum_{n_2\neq
A''}\frac{(\hat{\Sigma}(E_{A''}))_{A''n_2}(\gamma_0V_{\rm{PNC}})_{n_2A'}}{E_{n_2}-E_{A''}} \, ,
\end{align}
together with  $E_{A'}=E_{A''}$ in the denominator. 
This represents the  correction to the wave function $A''$ in the PNC
matrix element.

In order to account for  vacuum-polarization insertions we simply
have to replace Eq.~(\ref{q84}) by
\begin{equation}
\label{q85}  
(\gamma_0V_{\rm{PNC}})_{\tilde{A}''A'}=\sum_{n_2\neq
A''}\frac{\left[(\hat{\Sigma}(E_{A''}))_{A''n_2}+(V_U)_{A''n_2}\right](\gamma_0V_{\rm{PNC}})_{n_2A'}}{E_{n_2}-E_{A''}}
\, .
\end{equation}
A vacuum-polarization correction of this kind has been calculated numerically
in~\cite{qz+22}.

Now we turn to the evaluation of the SE insertions in the lower
electron propagator in Fig.~10 (see Fig.~12). The corresponding
scattering amplitude is
%\begin{multline}
\begin{eqnarray}
\label{q86} 
U_{\rm{sc}}=-\sum_{n_1}\frac{(U^*_{\omega})_{An_1}}{E_{n_1}-E_A-\omega} %\\ 
%\times 
\sum_{n_2n_3} \,\frac{(\gamma_0V_{\rm{PNC}})_{n_1n_2}
(\hat{\Sigma}(E_A+\omega))_{n_2n_3}(U_{\omega'})_{n_3A}}{(E_{n_2}-E_A-\omega)
(E_{n_3}-E_A-\omega)}\, .\nonumber\\
%\end{multline}
\end{eqnarray}
In this case the dominant state is  $n_1=A''$ and the resonant states
are $n_2=n_3=A'$, respectively. The expression for the emission amplitude 
can now be expressed as
\begin{equation}
\label{q87}  
U_{\rm{em}}=-\frac{(U^*_{\omega})_{AA''}}{E_{A''}-E_A-\omega}
\,\frac{(\gamma_0V_{\rm{PNC}})_{A''A'}}{E_{A'}-E_A-\omega}
\,\frac{(\hat{\Sigma}(E_A+\omega))_{A'A'}}{(E_{A'}-E_A-\omega)}\, .
\end{equation}
 After resummation of SE insertions for the resonant state $A'$
to all orders, we obtain
\begin{multline}
\label{q88}
%\begin{array}{l}
U_{\rm{em}} = \frac{(U^*_{\omega})_{AA''}}{\tilde{E}_{A''}-E_A-\omega}
\,\frac{(\gamma_0V_{\rm{PNC}})_{A''A'}}{\tilde{E}_{A'}-E_A-\omega}
= \frac{(U^*_{\omega})_{AA''}}{\tilde{E}_{A''}-\tilde{E}_{A'}}
\,\frac{(\gamma_0V_{\rm{PNC}})_{A''A'}}{\tilde{E}_{A'}-E_A-\omega}
\, .
%\end{array}
\end{multline}
Here we have used the resonance condition $\omega=\tilde{E}_{A'}-E_A$.

Again, the summation of all  SE corrections in
Fig.~11 is supposed to be already performed. Moreover, we will assume that VP
contributions are included in both  Fig.~11 and Fig.~12, respectively. Accordingly, we can write 
\begin{equation}
\label{q89}  
U_{\rm{em}}=\frac{(U^*_{\omega})_{AA''}}{\tilde{E}_{A'}-E_A-\omega}
\,\frac{(\gamma_0V_{\rm{PNC}})_{A''A'}}{\Delta E_L} \, ,
\end{equation}
together with  $\Delta E_L=L_{A''}-L_{A'}$. Eq.~(\ref{q89}) holds for the
degenerated case, when $E_{A'}=E_{A''}$.

Setting $n_2\neq A'$ in Fig.~12 we obtain the correction to the
PNC matrix element (respectively to the wave function of the state $A''$):
\begin{equation}
\label{q90}  
U_{\rm{em}}=\frac{(U^*_{\omega})_{AA''}}{\tilde{E}_{A'}-E_A-\omega}
\,\frac{(\gamma_0V_{\rm{PNC}})_{A''\tilde{A}'}}{\Delta E_L} \, ,
\end{equation}
where
\begin{equation}
\label{q91}  
(\gamma_0V_{\rm{PNC}})_{A''\tilde{A}'}=\sum_{n_2\neq A'}\frac
{(\gamma_0V_{\rm{PNC}})_{A''n_2}\left[(\hat{\Sigma}(E_{A''}))_{n_2A'}+
(V_U)_{n_2A'}\right]}
{E_{n_2}-E_{A'}}
\end{equation}
A similar replacement $E_{A''}-E_A-\omega\rightarrow\Delta E_L$  can
be performed in Eq.~(\ref{q83}).

Graphs with $V_{\rm{PNC}}$ inserted in the outer electron lines in
Fig.~10 are less important since they do not contain  dominant
terms with small denominators.

\section{Electromagnetic vertex correction to the weak interaction
mixed amplitude}
\label{sec14}

 The weak interaction
amplitude with one-loop SE corrections at the PNC vertex is described by the Feynman diagram in Fig.~13. The corresponding 
expression for this correction,  denoted  as
$U^{\rm{VW}}_{\rm{sc}}$, can be written as
\begin{equation}
\label{q92}  
U^{\rm{VW}}_{\rm{sc}}=-\sum_{n_1n_2}\frac
{(U^*_{\omega})_{An_1}(\Lambda^{(1)W}_{\mu}B^{\mu})_{n_1n_2}(U^*_{\omega'})_{n_2A}}
{(E_{n_1}-E_A-\omega)(E_{n_2}-E_A-\omega)}
\, .
\end{equation}
Here $\Lambda^{(1)W}_{\mu}$ is the weak interaction vertex that
differs from the pure electromagnetic vertex $\Lambda^{(1)}_{\mu}$ as 
discussed in section \ref{sec11} by changing the vector-coupling $\gamma_{\mu}$ that
enters in the expression for $\Lambda^{(1)}_{\mu}$ to the pseudo-vector matrix
$\gamma_5\gamma_{\mu}$.

A weak interaction potential $B_{\mu}$ according to
Eq.~(\ref{q77}) may be introduced  by
\begin{equation}
\label{q93}
B_{\mu}=-\delta_{\mu_0}\frac{G_F}{2\sqrt{2}}Q_W\rho_N(r) \, .
\end{equation}
Evaluating Eq. (\ref{q93}) within the resonance approximation implies $n_2=A'$ together with  
$n_1=A''$ as the dominant state.

Again we assume  that the resummation of radiative insertions in both
electron propagators in Fig.~10 is already performed:
\begin{equation}
\label{q94}   U^{\rm{VW}}_{\rm{em}}=\frac
{(U^*_{\omega})_{AA''}}{\tilde{E}_{A'}-E_A-\omega}\,\frac{(\Lambda^{(1)W}_{\mu}B^{\mu})_{A''A'}}
{\Delta E_L} \, .
\end{equation}
The vertex $\Lambda^{(1)W}_{\mu}$ is ultraviolet and infrared
divergent. The renormalization scheme can be taken over from the pure electromagnetic vertex:
\begin{align}
\label{q95}  
\Lambda^{(1)\rm{W(ren)}}_{\mu}&=\Lambda^{(1)W}_{\mu}(\not p', \not
p)-\gamma_{\mu}\Lambda^{(1)W} \, ,
\\
\label{q96}  
\gamma_{\mu}\Lambda^{(1)W} & \left.\equiv \Lambda^{(1)W}_{\mu}(\not
p', \not p)\right|_{\not p'=\not p=m} \, .
\end{align}
In view of  Eq.~(\ref{q65}) we can write
\begin{equation}
\label{q97}   \gamma_{\mu}\Lambda^{(1)W}=-4\pi
e^2\int\frac{d^4k}{(2\pi)^4}\gamma_{nu}\frac{\not
k}{k^2}\gamma_5\gamma_{\mu}\frac{\not
k}{k^2}\gamma^{\nu}\frac{1}{k^2}=\gamma_5\gamma_{\mu}\Lambda^{(1)}
\, .
\end{equation}
Similarly, the cancellation of the divergency in $U^{\rm{VW}}_{\rm{em}}$
occurs through the derivative corrections. Consider once more diagrams with
 SE insertions of the type as depicted in Fig.~11. Without
changing the results of section \ref{sec13} we can proceed with the first
term of the geometric progression in the same manner as it was done
in section \ref{sec11}. This yields
\begin{equation}
\label{q98}  
U^{0\rm{W}}_{\rm{em}}+U^{dW}_{\rm{em}}=\frac{(U^*_{\omega})_{AA''}}{\tilde{E}_{A'}-E_A-\omega}
\,\frac{(\gamma_0V_{\rm{PNC}})_{A''A'}}{\Delta%
E_L}\left(1+\Sigma_{A'}^{'(\rm{ren})}\right)\, .
\end{equation}
Combining Eq.~(\ref{q98}) with Eq.~(\ref{q94}) we have
\begin{multline}
\label{q99}
U^{0W}_{\rm{em}}+U^{\rm{VW}}_{\rm{em}}+U^{dW}_{\rm{em}} =  
\frac{(U^*_{\omega})_{AA''}}{\tilde{E}_{A'}-E_A-\omega}
\frac{1}{\Delta
 E_L} \\
 \times 
\left[(\gamma_0V_{\rm{PNC}})_{A''A'}+
(\Lambda^{(1)W(\rm{ren})}_{\mu}
B^{\mu})_{A''A'}  
+\Sigma_{A'}
^{'(\rm ren)}
(\gamma_{0}V_{\rm{PNC}})_{A''A'}\right]\, .
\end{multline}
Since $V_{\rm{PNC}}=\gamma_5B_0$ we find that the divergent part
of the second term in the square brackets in Eq.~(\ref{q99}) is
$-\Lambda^{(1)} (\gamma_0V_{\rm{PNC}})_{A''A'}$, while the divergent
part of the third term in  square brackets is $-\Sigma^{(1)'}
(\gamma_0V_{\rm{PNC}})_{A''A'}$. These divergent parts cancel due
to the Ward identity. The total expression can be rewritten in
symmetric form when performing the same manipulations with the first term
of the geometric progression in Fig.~12. As a result, we find
\begin{multline}
\label{q100}
U^{w}_{\rm em} = 
U^{0w}_{\rm{em}}+U^{VW}_{\rm{em}}+U^{dW}_{\rm{em}}    
 = \frac{(U^*_{\omega})_{AA''}}{{\tilde E}_{A'}-E_A-\omega}
\frac{1}{\Delta E_L}
\\
\times
\left[(\gamma_0 V_{\rm{PNC}})_{{A'}'A'}+
(\Lambda^{(1)W({\rm ren})}_{\mu}B^{\mu})_{A''A'}
+ \frac{  1}{  2}
(\Sigma_{A'}^{'\rm{(ren)}}+
\Sigma_{A''}^{'\rm{(ren)}}
)
\right]
(\gamma_{0} V_{\rm{PNC}} )_{A''A'}\, .
\end{multline}
In principle, there exists  also the VP electromagnetic vertex correction
depicted in Fig.~14. However, as it was shown in~\cite{qz+22}, this correction turns out to vanish within  Uehling approximation.

Finally, we can represent all  pure electromagnetic radiative
corrections to the PNC matrix element in the form
\begin{multline}
\label{q101} 
(\gamma_0\tilde{V}_{\rm{PNC}})_{A''A'}
=(\gamma_0V_{\rm{PNC}})_{A''A'}+(\gamma_0V_{\rm{PNC}})_{\tilde{A}''A'}+(\gamma_0V_{\rm{PNC}})_{A''\tilde{A}'}
 \\
 +(\Lambda^{(1)w(\rm ren)}_{\mu}B^{\mu})_{A''A'}+\frac{1}{2}\left(\Sigma_{A'}^{'(\rm ren)}
+\Sigma_{A''}^{'(\rm ren)}\right)(\gamma_0 V_{PNC})_{A''A'}\, .
\end{multline}
The VP parts of the corrections
$(\gamma_0V_{\rm{PNC}})_{A''\tilde{A}'}$ and
$(\gamma_0V_{\rm{PNC}})_{\tilde{A}''A'}$ have been calculated earlier
in~\cite{qz+22}.

Apart from  corrections to the weak interaction matrix element
there exist also corrections to the emission matrix element in the
weak interaction emission amplitude. They either result by setting
 $n_1\neq A''$, but $n_2=A''$ in Eq.~(\ref{q86}) or by inserting
radiative corrections into the upper external electron line in the diagram of
Fig.~12. In the latter case one may  turn over again to the
double photon resonance picture described in sections \ref{sec6}-\ref{sec10}.
The other corrections to the emission matrix element are given by the
Feynman graphs in Fig.~15. These corrections can be  treated in
the same manner as vertex corrections in the case of a pure electromagnetic
amplitude (section \ref{sec11}).

Note, that there exist unseparable  radiative effects which cannot
be adjusted to the weak interaction matrix element or to the emission
matrix element in the weak interaction emission amplitude. These
corrections are depicted in Fig.~16. However, they do not generate
small energy denominators of the type $(\Delta E_L)^{-1}$ and thus are
neglible compared to the corrections (\ref{q101}). Nevertheless,  they have
to be taken into account if for some reasons the small denominator
is absent.

\section{The interelectron interaction insertions in the electron
propagators}
\label{sec15}

In this section we begin to investigate the line profiles and
transition rates in two-electron ions. The same formalism will be
applicable to  few-electron ions as well.

For highly-charged ions the lowest-order radiative corrections and
the lowest-order interelectron interaction corrections (IIC) are
approximately of the same magnitude. While the first ones are of the order
$\alpha=e^2$  the second ones are of the order $1/Z$ which is
the same for  $\alpha Z\sim 1$. Since the lowest-order radiative
corrections and IIC are additive  the radiative
effects can be taken over from case of the one-electron ions (sections
\ref{sec2}-\ref{sec11}). What remains is to include the lowest order IIC, i.e.
one-photon exchange corrections.

The first-, second- and partly the third-order IIC to the energy
levels have been calculated within the framework of QED during the last
decade for two- and three-electron
ions~\cite{qz+33}-~\cite{qz+37}, \cite{qz+14}. In particular, the
line-profile approach was used for the evaluation of the reducible
parts of the two- and three-photon exchange corrections
in~\cite{qz+14},~\cite{qz+39},~\cite{qz+13}.

However, the interelectron interaction corrections to the
transition probabilities have not been studied so thoroughly. In this
and the following sections of our paper we present a full QED
approach to the evaluation of the IIC based on the line profile
theory. We will mainly concentrate on the first-order IIC, although
the generalization to higher orders is straightforward.

Consider the elastic photon scattering on a two-electron ion
with the first-order IIC taken into account. We assume that 
scattering occurs at the electron in the state $A$, while the
electron in the state $B$ plays the role of a spectator. This
process is depicted in Fig.~17. For the description of the
interelectron interaction we use the Coulomb gauge, distinguishing
the exchange by the Coulomb and Breit (transverse) photons. The
contribution of the graph in Fig.~17 to the scattering amplitude
is given by
\begin{equation}
\label{q102}  
U_{\rm{sc}}=\sum_{n_1}\frac{(U^*_{\omega})_{An_1}}{E_{n_1}-E_A-\omega}
\sum_{n_2}\frac{(V_{\rm{C}})_{n_1B;n_2B}(U^*_{\omega'})_{n_1A}}{E_{n_2}-E_A-\omega} \, ,
\end{equation}
where $V_{\rm{C}}$ is the Coulomb interaction
\begin{equation}
\label{q103}  
(V_{\rm{C}})_{A'B'AB}=\int\Psi^{+}_{A'}(\vec{x}_1)\Psi^{+}_{B'}(\vec{x}_2)\frac{e^2}{r_{12}}
\Psi_{A}(\vec{x}_1)\Psi_{B}(\vec{x}_2)\, d\,^3x_1d\,^3x_2\, .
\end{equation}
We adopt  the notation
\begin{equation}
\label{q104}  
(\hat{F})_{A'B';AB}\equiv(\hat{F})_{A'B'AB}-(\hat{F})_{B'A'AB} \, ,
\end{equation}
where $\hat{F}$ denotes an arbitrary two-electron operator. The second
term in Eq.~(\ref{q104}) takes into account the "exchange" part of
the interaction.

 As in earlier cases we employ the resonance condition $E_{A'}=E_A+\omega$ and
 set $n_1=n_2=A'$ leading to the correction to the emission
amplitude in  resonance approximation
\begin{equation}
\label{q105}  
U_{\rm{sc}}=\frac{(U^{*}_{\omega})_{AA'}}{E_{A'}-E_A-\omega}
\,\frac{(V_{\rm{C}})_{A'B;A'B}}{E_{A'}-E_A-\omega}\, .
\end{equation}
The next iteration with respect to the Coulomb interaction $V_{\rm{C}}$ follows from the
graph in Fig.~18. In this case the resonance
approximation implies that $n_1,n_4=A$ and $n_2,n_3\in
A',B$. The inclusion of all possible combinations leads to the
correction
\begin{equation}
\label{q106}  
U_{\rm{em}}=\frac{(U^*_{\omega})_{AA'}}{E_{A'}-E_A-\omega}
\,\frac{\left[(V_{\rm{C}})_{A'B;A'B}\right]^2}{(E_{A'}-E_A-\omega)^2}
\, .
\end{equation}
An all-order resummation of the iterations results in a geometric
progression and thus in an energy shift in the denominator
\begin{equation}
\label{q107}  
U_{\rm{em}}=\frac{(U^*_{\omega})_{AA'}}{E_{A'}+\Delta
E^{c1}_{A'B}-E_A-\omega} \, .
\end{equation}
Here $\Delta E^{c1}_{A'B}$ is the first-order Coulomb IIC:
\begin{equation}
\label{q108}   \Delta
E^{c1}_{A'B}=(V_{\rm{C}})_{A'B;A'B} \, .
\end{equation}
Exactly the same derivation can be repeated for the
Breit interaction leading to  the expression (see Fig.~19):
\begin{equation}
\label{q109}  
U_{\rm{em}}=\frac{(U^*_{\omega})_{AA'}}{E_{A'}+\Delta
E^{1}_{A'B}-E_A-\omega}\, ,
\end{equation}
where
\begin{align}
\label{q110}  
\Delta E^{1}_{A'B}&= \Delta
E^{C1}_{A'B}+\Delta E^{B1}_{A'B} ,
\\
\label{q111}  \Delta
E^{B1}_{A'B}&= (V_{\rm{B}})_{A'B;A'B} .
\end{align}
The matrix elements of the Breit interaction $V_{\rm{B}}$ are
defined as:
\begin{align}
\label{q112}  
(V_{\rm{B}})_{A'BA'B}&= -e^2\left(\frac{\alpha_{i_1}\alpha_{i_2}}{r_{12}}\right)_{A'BA'B}
\, ,
\\
\label{q113}  (V_B)_{BA'A'B}&= e^2 \left(
\frac{\alpha_{i_1}\alpha_{i_2}-1}{r_{12}}e^{i|E_{A'}-E_B|r_{12}}
\right)_{BA'A'B} \, .
\end{align}

Here $\alpha_{i_1}, \alpha_{i_2}$ denote the Dirac matrices which act on
 one-electron wave functions depending on spatial variables $\vec{x}_1,
\vec{x}_2$ and $i=1,2,3$ .

Moreover, we can perform simultaneously the resummation of the
radiative insertions in the electron propagator. 
According to the
derivations presented in section \ref{sec3}, this yields the result
\begin{equation}
\label{q114} 
U_{\rm{em}}=\frac{(U^*_{\omega})_{AA'}}{\tilde{E}_{A'}+\Delta
E^1_{A'B}-E_A-\omega} \, .
\end{equation}
The emission line profile can now be expressed, as described
in section \ref{sec4}.

Following the derivations of section \ref{sec5} we account for the
higher-order IIC as well. For example the second-order Coulomb-Coulomb IIC
can be obtained from Fig.~18 for $n_1,n_4=A'$ but
$n_2,n_3\bar{\in}A',B$. The part of this graph containing two Coulomb
interactions can  now be considered as a complicated insertion into
the electron line corresponding to the state $A$. Repeating these
insertions and summing up the resulting geometric progression we shall
 obtain an additional shift in the denominator that will
represent exactly the second-order Coulomb-Coulomb IIC. Similar manipulations
 can be performed 
 for the second-order Coulomb-Breit and Breit-Breit IIC. In
the two latter cases  reducible contributions also arise due
the dependence of the IIC corrections on the photon frequency
$\omega$~\cite{qz+14},~\cite{qz+39},~\cite{qz+40}.

These reducible contributions, i.e., the so-called reference-state
contributions or derivative contributions can be obtained in the
same manner as the reducible contributions discussed in section \ref{sec5}.

The derivation of the IIC correction to the emission amplitude follows
 to the derivation of the radiative correction presented in section
\ref{sec6a}. Assuming now that in Fig.~17 and Fig.~19 $n_1\neq A'$  but
$n_2=A'$ and that all previous  insertions are also performed as well,
 we obtain
the generalization of Eq.~(\ref{q114})
\begin{equation}
\label{q115} 
U_{\rm{em}}=\frac{(U^*_{\omega})_{A\tilde{A}'}}{\tilde{E}_{A'}+\Delta
E^1_{A'B}-E_A-\omega}\, .
\end{equation}
Here
\begin{equation}
\label{q116} 
(U^*_{\omega})_{A\tilde{A}'}=\sum_{n_1\neq A'}\frac
{(U^*_{\omega})_{An_1}\left[(V_{\rm{C}})_{Bn_1;BA'}+(V_{\rm{B}})_{Bn_1;BA'}\right]}
{E_{n_1}-E_{A'}}
\end{equation}
is the first-order IIC to the emission matrix element
(correction to the wave function of the state $A'$).

To obtain the shift of the energy $E_A$ in the denominator of
Eq.~(\ref{q115}), as well as the correction to the wave function
$\Psi_A$ in the emission matrix element we have to turn over again to
the double resonance photon scattering picture as described in
section \ref{sec11}. The following section will be devoted to this issue.

\section{ Double photon scattering on two-electron ions}
\label{sec16}

The double photon resonant scattering process described in section
\ref{sec6} can be applied without any modifications to two-electron ions.
The first-order IIC insertions into the central electron
propagator in Fig.~5 will yield as was shown in section \ref{sec7} for
the radiative insertions the same result that was already
obtained with the single photon scattering: the shift of the
energy $E_{A'}$ in the Lorentz denominator and the correction to
the wave function $\Psi_{A'}$ in the emission matrix element.
Instead of repeating this derivation  we start
with the IIC insertions into the upper electron propagator in
Fig.~5. These insertions to first order are shown in Fig.~20.
Applying similar steps as described in  section \ref{sec8}, we
derive an expression for the double photon emission amplitude
\begin{equation}
\label{q117}  U_{\rm{em}}=\frac
{(U^*_{\omega_0})_{A_0A}(U^*_{\omega})_{AA'}}{(\tilde{E}_A+\Delta
E^1_{AB}-E_{A_0}-\omega)(\tilde{E}_{A'}+\Delta
E^1_{A'B}-E_{A_0}-\omega_0-\omega)}\, .
\end{equation}
For obtaining Eq.~(\ref{q117}) we have set $n_1=n_2=A, n_3=A',
n_4=A$ in the diagrams Fig.~20 a)-d). We can also generalize Eqs.~(\ref{q56}),
(\ref{q57}) for the Lorentz line profile both for the excited and
for the ground state $A$. The only difference to Eqs.~(\ref{q56}),
(\ref{q57}) is that now the frequency $\tilde{\omega}_{A'A}$
 also accounts for  the IIC
\begin{equation}
\label{q118} 
\tilde{\omega}_{A'A(B)}=\tilde{E}_A+\Delta
E_{AB}-\tilde{E}_{A'}-\Delta E_{A'B}\, .
\end{equation}
Such interelectronic interaction corrections can be of any order
\begin{equation}
\label{q119}  \Delta E_{AB}=\sum_{i=1}^{\infty}\Delta
E^i_{AB}
\end{equation}
and the same holds true for $\Delta E_{A'B}$.

Accordingly we can replace $\tilde{\omega}_{A'A(B)}$ by its experimental
value $\omega^{\rm{exp}}_{A'A(B)}$. This is especially important
for two-electron ions, where the energy differences between the
levels within a multiplet can be deduced experimentally more
accurately, than from theory.

Finally, we can write down the correction to the emission matrix
element (to the wave function of the state $A$), following 
derivations presented in section \ref{sec10}
\begin{equation}
\label{q120} 
U_{\rm{em}}=\frac{(U^*_{\omega_0})(U^*_{\omega})_{\tilde{A}A'}}
{(\tilde{E}_A+\Delta
E_{AB}-E_{A_0}-\omega_0)(\tilde{E}_{A'}+\Delta
E_{A'B}-E_{A_0}-\omega_0-\omega)} \, ,
\end{equation}
where
\begin{equation}
\label{q121} 
(U^*_{\omega})_{\tilde{A}A'}=\sum_{n_2\neq A}\frac
{\left[(V_{\rm{C}})_{BA;Bn_2}+(V_{\rm{B}})_{BA;Bn_2}\right](U^*_{\omega})_{n_2A'}}
{E_{n_2}-E_A} \, .
\end{equation}
For deriving Eq.~(\ref{q121}) we have set $n_1=A,
n_3=A', n_4=A,$ but now with $n_2\neq A$ in Figs.~20 a)-d).

Again we can write down all the corrections to the transition
probability $A'B\rightarrow AB$ (or the partial width
$\Gamma_{AB;A'B}$) in the form of Eq.~(\ref{q61}).

The first-order IIC to the transition probability can be always
represented by the sum
\begin{equation}
\label{q122} 
\Gamma_{AB,A'B}=\Gamma_{AA'}+\Gamma_{\tilde{A}A'}+\Gamma_{A\tilde{A}'}
\end{equation}

where $\Gamma^{(B)}_{\tilde{A}(B)A'}$ and
$\Gamma^C_{A\tilde{A}'(B)}$ represent one-electron partial widths with
improved amplitudes given by Eqs.~(\ref{q116}) and (\ref{q121}), respectively.

\section{The interelectron interaction corrections to the weak
interaction amplitude}
\label{sec17}

To evaluate the IIC corrections to the weak interaction amplitude
we can take the  sections \ref{sec12}-\ref{sec13} as a guide-line. The generalization
of the diagrams in Fig.~10 to two-electron ions with IIC insertions into the
electron propagators is depicted in Figs.~21 and 22.

Consider first the IIC in Fig.~21. Here we take 
$n_3=E_{A'}$ (resonant case) and $n_1=n_2=E_{A''}$ (dominant
state). Then the summation of IIC insertions to all orders will
result in (compare with Eq.~(\ref{q82}))
\begin{equation}
\label{q123}  
U_{\rm{em}}=\frac{(U^*_{\omega})_{AA''}}{\tilde{E}_{A''}+\Delta
E_{A''B}-E_A-\omega}
\,\frac{(\gamma_0V_{\rm{PNC}})_{A''A'}}{\tilde{E}_{A'}-E_A-\omega} \, .
\end{equation}
We assume that the radiative insertions are already performed. If
$n_2\neq A''$ we obtain the correction to the PNC matrix element
(to the wave function $\Psi_{A''}$)
\begin{equation}
\label{q124}  
U_{\rm{em}}=\frac{(U^*_{\omega})_{AA''}}{\tilde{E}_{A''}+\Delta
E_{A''B}-E_A-\omega}
\frac{(\gamma_0V_{\rm{PNC}})_{\tilde{A}''A'}}{\tilde{E}_{A'}-E_A-\omega} \, ,
\end{equation}
where
\begin{equation}
\label{q125}  
(\gamma_0V_{\rm{PNC}})_{\tilde{A}''A'}=\sum_{n_2\neq A''}\frac
{\left[(V_{\rm{C}})_{BA'';Bn_2}+(V_{\rm{B}})_{BA'';Bn_2}\right](\gamma_0V_{\rm{PNC}})_{n_2A'}}
{E_{n_2}-E_{A''}}\, .
\end{equation}
For the IIC in Fig.~22 we have to choose the dominant state
$n_1=A''$ and the resonant states $n_1=n_2=A'$. Then the summation
of IIC insertions to all orders yields (assuming again that the
summation in Fig.~21 is already performed)
\begin{equation}
\label{q126}  
U_{\rm{em}}=\frac{(U^*_{\omega})_{AA''}}{\tilde{E}_{A''}+\Delta
E_{A''B}-E_A-\omega}
\,\frac{(\gamma_0V_{\rm{PNC}})_{A''A'}}{\tilde{E}_{A'}+\Delta
E_{A'B} -E_A-\omega} \, .
\end{equation}
As in section \ref{sec13} we employ the resonance condition
$\omega=\tilde{E}_{A'}+\Delta E_{A'B}-E_A$ obtained from the
second denominator in Eq.~(\ref{q126}) and insert it into the first
denominator:
\begin{equation}
\label{q127}  
U_{\rm{em}}=\frac{(U^*_{\omega})_{AA''}}{\tilde{E}_{A''}+\Delta
E_{A''B}-\tilde{E}_{A'}-\Delta E_{A'B}}
\,\frac{(\gamma_0V_{\rm{PNC}})_{A''A'}}{\tilde{E}_{A'}+\Delta
E_{A'B} -E_A-\omega}\, .
\end{equation}
 The meaning of "dominant" state for
two-electron ions needs to be clarified. These states should now  be specified in a different
manner compared to one-electron ions. The most advantageous situation
for the observation of the effects in two-electron ions occurs
when two levels belonging to the same multiplet but possessing
opposite parity appear to be nearly degenerated. Unlike in the case
of one-electron ions in two-electron ions this can happen only
accidentally. The standard example is the crossing of the
$1s2s{}^1S_0$ and $1s2p{}^3P_0$ levels that occurs for $Z$ = 6
(carbon), $Z$ = 64 (gadolinium) and $Z$ = 92 (uranium) He-like
ions~\cite{qz+15}-~\cite{qz+19}),~\cite{qz+21}.

Introducing the frequency
\begin{equation}
\label{q128}  
\tilde{\omega}_{A''A'(B)}=\tilde{E}_{A''}+\Delta
E_{A''B}-\tilde{E}_{A'}-\Delta E_{A'B}
\end{equation}
we may replace it then by the experimental value
$\tilde{\omega}^{\rm{exp}}_{A''A'(B)}$. The exact theoretical
value for the splitting $\tilde{\omega}_{A''A'(B)}$ between the
nearly degenerated levels of the two-electron ions can be also
used.  Now the energy difference $\tilde{\omega}_{A''A'(B)}$ plays
the role of the small denominator $\Delta E_L$ and we can write down the
analogue of Eq.~(\ref{q89}) in the form
\begin{equation}
\label{q129}  
U_{\rm{em}}=\frac{(U^*_{\omega})_{AA''}}{\tilde{E}_{A'(B)}-E_A-\omega}
\,\frac{(\gamma_0V_{\rm{PNC}})_{A''A'}}{\Delta E_L}
\end{equation}
where $\Delta E_L=\tilde{\omega}_{A''A'(B)}$ and
\begin{equation}
\label{q130}   \tilde{E}_{A'(B)}=\tilde{E}_{A'}+\Delta
E_{A'B}\, .
\end{equation}
Setting now  $n_2\neq A'$ in Fig.~22 we obtain the correction to the
PNC matrix element (to the wave function in the state $A'$)
\begin{equation}
\label{q131}  
U_{\rm{em}}=\frac{(U^*_{\omega})_{AA''}}{\tilde{E}_{A'(B)}-E_A-\omega}
\,\frac{(\gamma_0V_{\rm{PNC}})_{A''\tilde{A}'}}{\Delta E_L} \, ,
\end{equation}
where
\begin{equation}
\label{q132}  
(\gamma_0V_{\rm{PNC}})_{A''\tilde{A}'}=\sum_{n_2\neq A'}\frac
{(\gamma_0V_{\rm{PNC}})_{A''n_2}\left[(V_{\rm{C}})_{Bn_2;BA'}+(V_{\rm{B}})_{Bn_2;BA'}\right]}
{E_{n_2}-E_{A'}}\, .
\end{equation}
The similar replacement
$\tilde{E}_{A''(B)}-E_A-\omega\rightarrow\Delta E_L$ can be made
in Eq.~(\ref{q124}).

Again we do not consider the graphs with $V_{\rm{PNC}}$ inserted
in the outer electron lines, since they do not contain any small
denominators $\Delta E_L$.

Finally, we end up with  an expression for the PNC matrix element with all the IIC
taken into account
\begin{equation}
\label{q133}  
(\gamma_0V_{\rm{PNC}})_{A''A'}=
(\gamma_0V_{\rm{PNC}})_{A''A'}+(\gamma_0V_{\rm{PNC}})_{A''\tilde{A}'}+(\gamma_0V_{\rm{PNC}})_{\tilde{A}''A'}
\, ,
\end{equation}
where the corrected matrix elements
$(\gamma_0V_{\rm{PNC}})_{A''\tilde{A}'}$ and
$(\gamma_0V_{\rm{PNC}})_{\tilde{A}''A'}$ are defined by
Eqs.~(\ref{q132}) and (\ref{q125}), respectively.

\section{The interelectron interaction corrections in the
Dirac-Hartree-Fock approximation}
\label{sec18}

The QED theory of highly-charged ions, based on the
Dirac-Hartree-Fock (DHF) taken as zero-order approximation has been  considered
in~\cite{qx+1}. Due to the energy dependence in the Breit-interaction 
matrix elements (see Eq.~(\ref{q113})), the Breit-interaction
 cannot be included exactly in the zero-order DHF
scheme. Therefore, we will consider here the DHF approximation restricting to
 the Coulomb interaction only and  leaving the Breit interaction to be
accounted perturbatively. Nevertheless, we may emphazise that the DHF method together 
with an approximate inclusion of the Breit interaction is widely used
and leads to very accurate results even for high-$Z$ ions (see, for
example~\cite{qz+1}).

In DHF approximation the one-electron equation (\ref{q0}) should
be replaced by 
\begin{equation}
\label{q134}
(\not{p}-m+\gamma_{0}eV+\gamma_{0}eV_{\rm HF})\Psi=0
\end{equation}
where the operator $V_{\rm{HF}}$ is defined as
\begin{multline}
\label{q135} 
V_{\rm{HF}}(\vec{r}\,)f(\vec{r}\,)=
\sum^N_{m=1}\int \,d\,^3 r'\, \Psi^+_m({\vec r}\,')
V_{\rm{C}}(\vec{r},{\vec r}\,')
\Psi_m({\vec r}\,')
f(\vec{r}\,)
 \\
 -
\sum^N_{m=1}
\int\, d\,^3r \,\Psi^+_m({\vec r}\,')
V_{\rm C }
(\vec{r},\vec{r}\,')
f({\vec r}\,')
\Psi_m(\vec{r}\,)\, .
\end{multline} 
In Eq.~(\ref{q135}) $\Psi_m(\vec{r}\,)$ are the solutions of the DHF
equation with eigenvalues $E_m, f(\vec{r}\,)$ is an arbitrary
4-component spinor function and the summation is extended to all occupied
states of the $N$-electron atom or ion.

For highly-charged ions we can consider the term
$\gamma_0V_{\rm{HF}}$ in Eq.~(\ref{q134}) as a perturbation. Then
the first-order correction to the Coulomb wave function $\Psi_A$
(the solution of Eq.~(\ref{q0})) will read
\begin{equation}
\label{q136}  \Psi^1_A(\vec{r}\,)=\sum_{n\neq
A}\frac{(V_{\rm{HF}})_{nA}}{E_n-E_A}\Psi_n(\vec{r}\,) \, .
\end{equation}

Using the definition of the operator $V_{\rm{HF}}$, we obtain´
\begin{equation}
\label{q137} 
\Psi^1_A(\vec{r}\,)=\sum^N_{m=1}\sum_{n\neq
A}\frac{(V_C)_{nm;Am}\Psi_n(\vec{r}\,)}{E_n-E_A} \, .
\end{equation}
In case of two-electron ions the sum over $m$ in Eq.~(\ref{q137})
reduces to one term $m=B$. Then, we obtain the corrections
$(U^*_{\omega})_{A\tilde{A}'}$ and $(U^*_{\omega})_{\tilde{A}A'}$
to the emission matrix element (Eqs.~(\ref{q116}) and (\ref{q121})
 only with the Coulomb potential $V_{\rm{C}}$) if we replace the
corrected wave functions $\Psi_{\tilde{A}}$ and  $\Psi_{\tilde{A}'}$
by the expression (\ref{q137}). This means that when 
evaluating the emission matrix elements with DHF functions we will
automatically take into account the corrections Eq.~(\ref{q116})
and Eq.~(\ref{q121}) with the Coulomb operator. The DHF functions
 include also  higher-order Coulomb corrections which are negligible
for high-$Z$ values. The same holds true for the corrections to the PNC matrix
elements as given  by  Eqs.~(\ref{q125}) and (\ref{q132}), respectively.

\section{Conclusions}
\label{sec19}

Summarizing the results obtained in the present paper we can state that a
most general QED approach has been elaborated for the evaluation of
various spectroscopical properties of highly-charged ions: energy
level shifts, transition rates and line profiles. 
The physical
process of  photon scattering on atomic electrons provides
a proper basis for  rigorous QED evaluations of all  quantities 
being of current interest in studies of highly charged ions.

The resonance approximation has been proven to be a powerful tool
for obtaining nonperturbative results such as the energy shifts in
the Lorentz denominators. This is a most physical way to determine
the atomic energy shifts.

Rigorous QED expressions for all generic radiative corrections to the
photon emission process are presented including simultaneously the
radiative shift of the Lorentz denominator. Thus a full QED
treatment of the Lorentz profile is achieved. The weak
interaction-induced parity violating amplitudes are also treated within
 this approach and a closed expression for  radiative
corrections to the PNC matrix element has been derived. This
allows for future accurate calculations of the PNC effect in
highly charged ions. Furthermore, the  theory is extended also to few-electron
ions.

 QED and PNC corrections to the emission amplitude are
presented for the most interesting case of single-excited states
in two-electron ions. The renormalization procedure for both pure
QED radiative corrections and  QED corrections in the presence
of a parity-violating effective potential is discussed and the
cancellation of the infrared divergencies is verified.

The general QED approach and the formulas for the various QED
corrections derived in the paper provide a firm theoretical basis
for numerical calculations in the near future.

\section*{Acknowledgments}

L.N. is grateful to the Technical University of Dresden
and to the Max-Planck-Institut  f\"ur Physik Komplexer Systeme for 
the hospitality during his visit in 2001. 
The work of ~L.~L., ~A.~P., and ~A.~S. was supported by the RFBR grant 
$N^{o}$ 99-02-18526 and by Minobrazovanije grant $N^{o}$ E00-3.1.-7. 
G.P. and G.S. acknowledge support by BMBF, DFG and by
GSI(Darmstadt).

\clearpage

\begin{figure}

\includegraphics[bb = 210 400 380 650,scale=1.4]{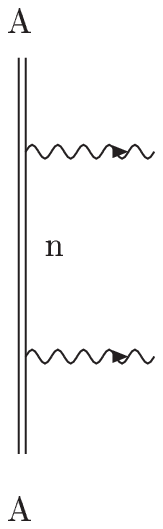}
%{{\includegraphics[bb = 300 375 380 640,scale=1.4]{figure1.ps}}

\caption{ Feynman graph describing the elastic photon scattering on 
an atomic electron. The double solid line denotes the bound electron
in the field of the nucleus. The wavy lines with the arrows denote
the photon emission and absorption. The indices A, n correspond to
the initial (final) and intermediate states.}
\label{fig1} 
\end{figure}
\clearpage
\begin{figure}
 
\includegraphics[bb = 210 400 380 650,scale=1.4]{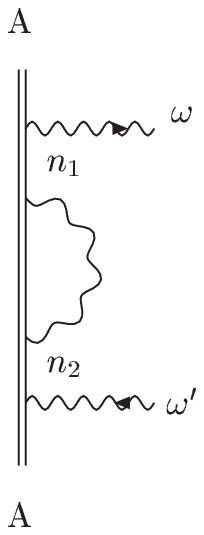}

\caption{ Feynman graph  corresponding to the electron self-energy
insertions into the electron propagator in graph Fig.~1. The
notations are the same as in Fig.~1. }
\label{fig2} 
\end{figure}
\clearpage
\begin{figure}

\includegraphics[bb = 210 400 410 650,scale=1.4]{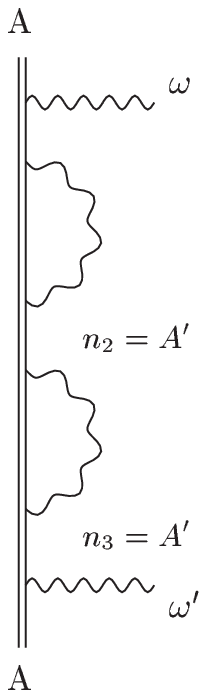}

\caption{  Feynman graph corresponding to double self-energy insertions
in the resonance approximation. }
\label{fig3} 
\end{figure}
\clearpage
\begin{figure}

\includegraphics[bb = 180 350 445 650,scale=1.4]{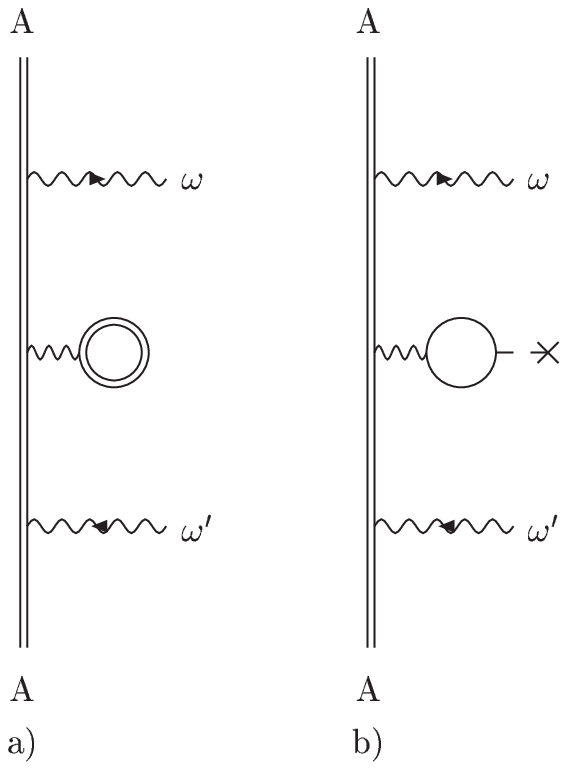}
\caption{Vacuum polarization insertions in the electron propagator. Part  a)
corresponds to the exact expression and part b) describes the Uehling 
approximation. The ordinary solid line in Fig.~4b denotes the free
electron propagation, the dashed line with the cross at the end denotes the
interaction with the nuclear potential.}  
\label{fig4} 
\end{figure}
\clearpage
\begin{figure}

\includegraphics[bb = 210 400 435 700,scale=1.4]{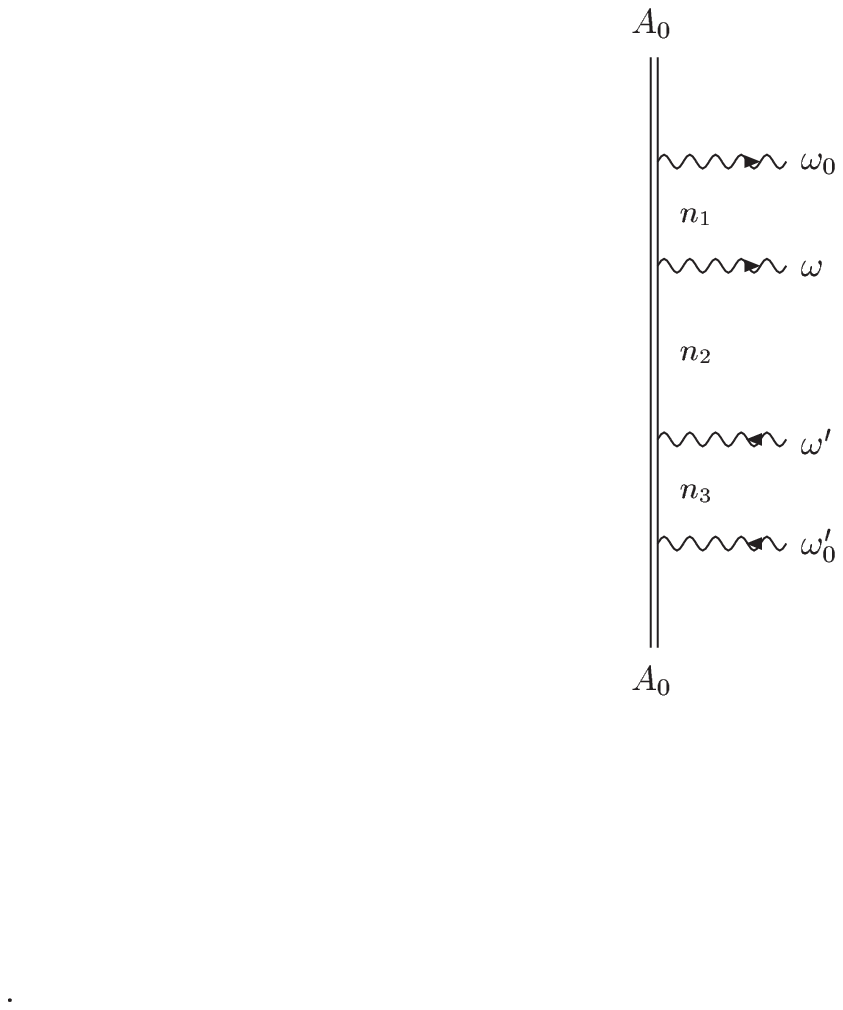} 
\caption{  Feynman graph describing the elastic scattering of two
photons on an  atomic electron in the state $A_0$.} 
\label{fig5} 
\end{figure}
\clearpage
\begin{figure}
 
\includegraphics[bb = 210 400 445 700,scale=1.4]{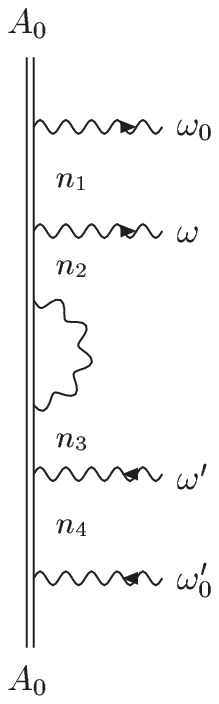} 
\caption{ Feynman graph corresponding to the electron self-energy
insertion in the central electron propagator in Fig.~5.}
\label{fig6} 
\end{figure}
\clearpage
\begin{figure} 

\includegraphics[bb = 210 400 445 700,scale=1.4]{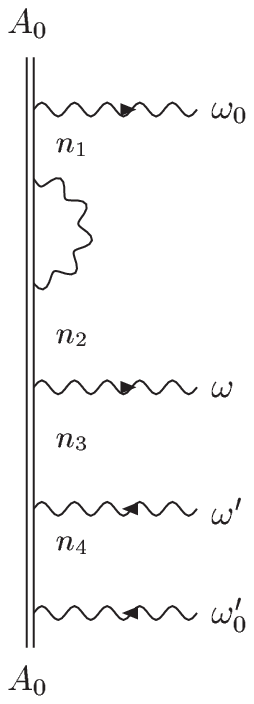}

\caption{ Feynman graph corresponding to self-energy insertions in the
upper electron propagator in Fig.~5.}     
\label{fig7} 
\end{figure}
\clearpage
\begin{figure}
  
\includegraphics[bb = 210 430 445 700,scale=1.6]{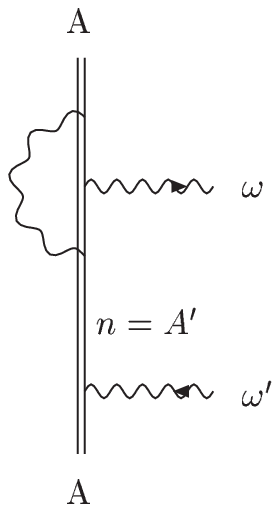}
 
\caption{The self-energy vertex correction to the transition amplitude
$A'\rightarrow A$ in the resonance approximation.}   
\label{fig8} 
\end{figure}
\clearpage
\begin{figure}
  
\includegraphics[bb = 190 400 445 700,scale=1.4]{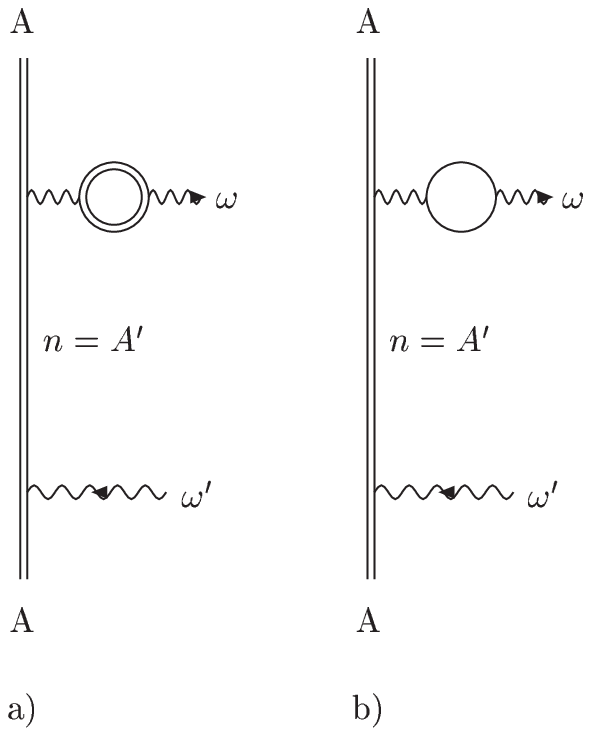}
 
\caption{The vacuum polarization vertex correction to the transition
amplitude $A'\rightarrow A$ in the resonance approximation: a)
exact expression, b) Uehling approximation.}
\label{fig9}
\end{figure}
\clearpage
\begin{figure}
\includegraphics[bb = 210 520 450 700,scale=1.6]{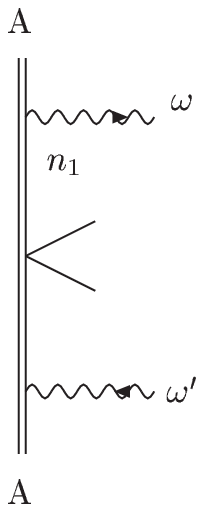} 
\caption{ Feynman graph that describes the elastic photon scattering on
an atomic electron in the presence of weak interaction. The
effective potential of  weak interaction between the electron
and the nucleus is denoted by a "sea-gull".}
\label{fig10}
\end{figure}
\clearpage

\begin{figure}
\includegraphics[bb = 200 400 420 700,scale=1.4]{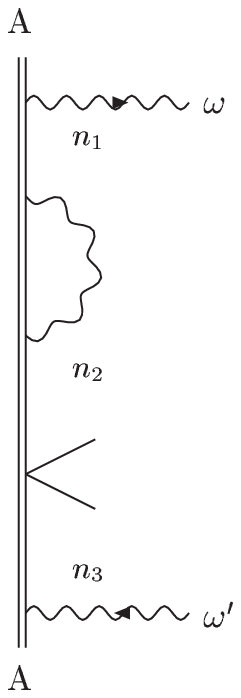}
\caption{The self-energy insertion in the upper electron propagator in
Fig.~10.}    
\label{fig11}
\end{figure}
\clearpage
\begin{figure}
 
\includegraphics[bb = 210 400 400 700,scale=1.4]{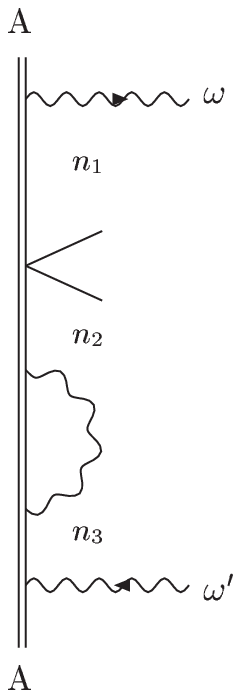} 
\caption{The self-energy insertions in the lower electron propagator in
Fig.~10}
\label{fig12}
\end{figure}
\clearpage
\begin{figure}[b] 
\includegraphics[bb = 210 400 400 700,scale=1.4]{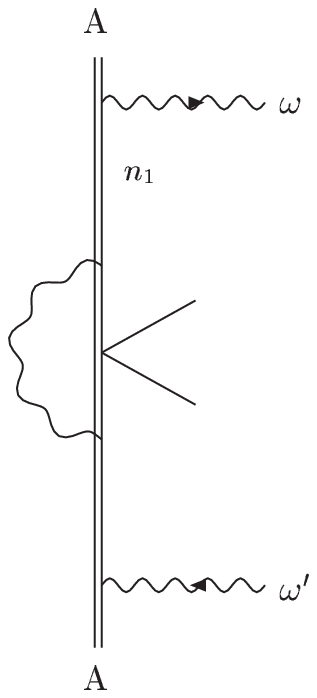} 
\caption{ Feynman graph describing the self-energy electromagnetic
vertex correction to the weak amplitude in  Fig.~10.}
\label{fig13}
\end{figure}
\clearpage
\begin{figure}
 
\includegraphics[bb = 170 390 410 700,scale=1.2]{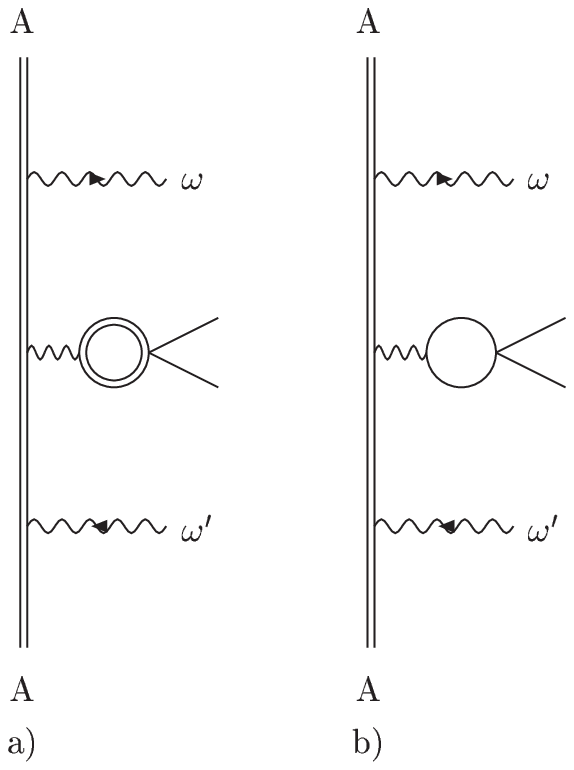}
%{\includegraphics[scale=1.4]{figure14.ps}} 
\caption{ Feynman graphs describing the vacuum polarization
electromagnetic vertex correction to the weak amplitude in Fig.~10:
a) exact expression b) Uehling approximation.}
\label{fig14}
\end{figure}
\clearpage
\begin{figure}

\includegraphics[bb = 130 380 520 700,scale=1.2]{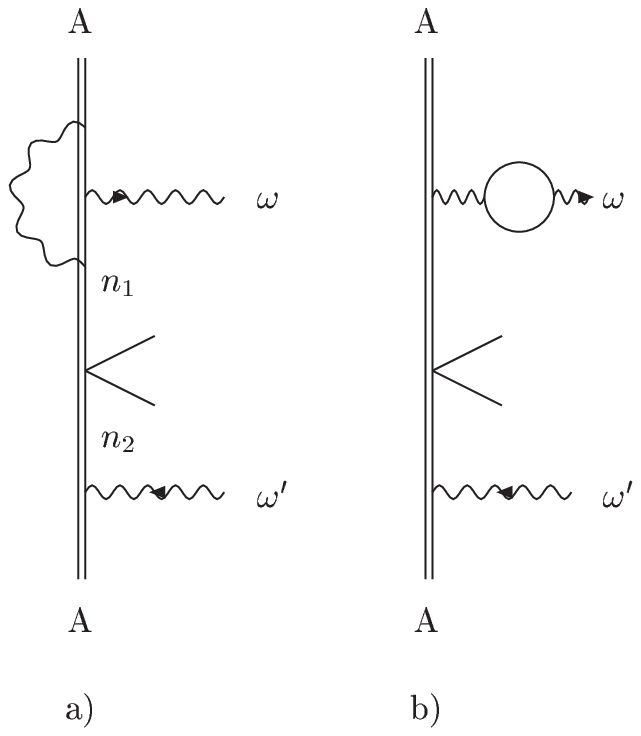} 
\caption{Vertex corrections to the emission matrix element in the weak
emission amplitude a) self-energy vertex correction b) vacuum
polarization vertex correction in  Uehling approximation.}
\label{fig15}
\end{figure}
\clearpage

\begin{figure}
\includegraphics[bb = 150 400 420 700,scale=1.2]{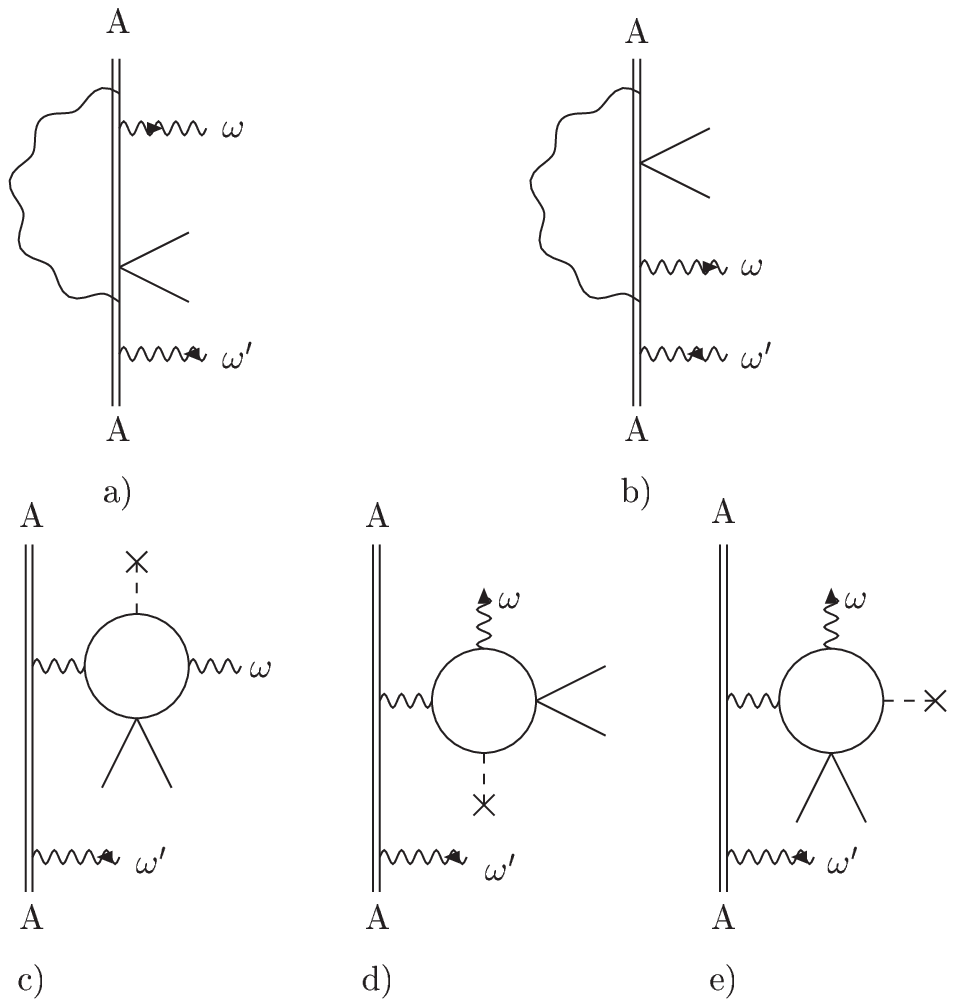}
\caption{Unseparable radiative corrections to the weak emission amplitude.
Figs.~16 a), b) correspond to self-energy corrections, Figs.~16
c)-e) correspond to vacuum polarization corrections in  Uehling
approximation.} 
\label{fig16}
\end{figure}
\clearpage
\begin{figure}

\includegraphics[bb = 150 550 600 700,scale=1.1]{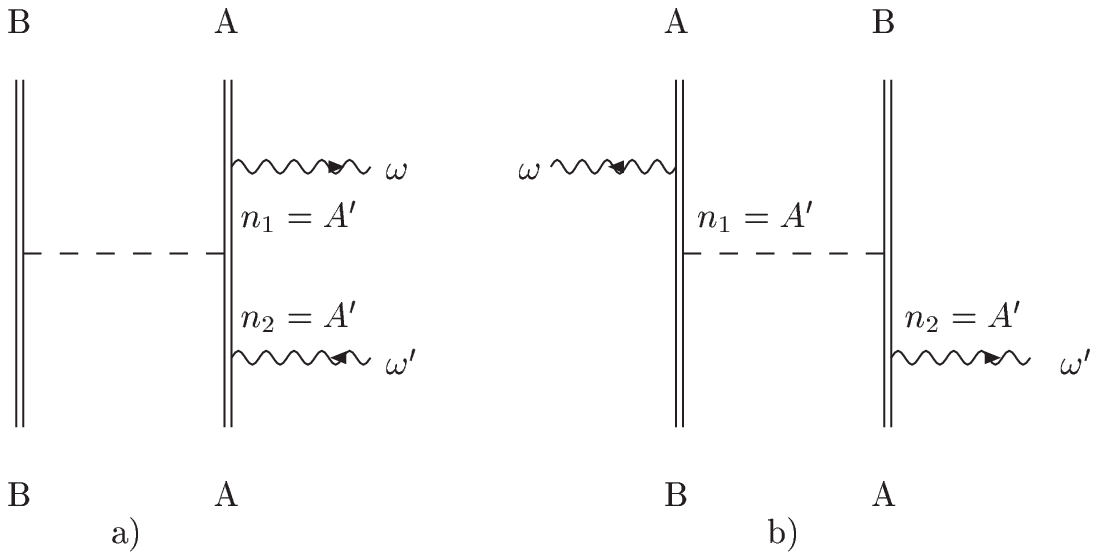} 
\caption{The Feynman graphs that describe the elastic photon scattering on
a two-electron ion in  resonance approximation with 
first-order Coulomb interelectron interaction corrections taken
into account. Here $A, B$ denote one-electron states in the
two-electron ion, $A'$ is the resonant state. The dashed line
corresponds to the Coulomb photon. The graph Fig.~17a) describes
the "direct" term in the amplitude and the graph Fig.~17b)
corresponds to the ''exchange'' term. The contributions of the
"exchange" graphs always enter with an additional minus sign which traces back
  to the antisymmetrization of the two-electron wave
function.}
\label{fig17}
\end{figure}
\clearpage
\begin{figure}

\includegraphics[bb = 150 400 400 800,scale=1.1]{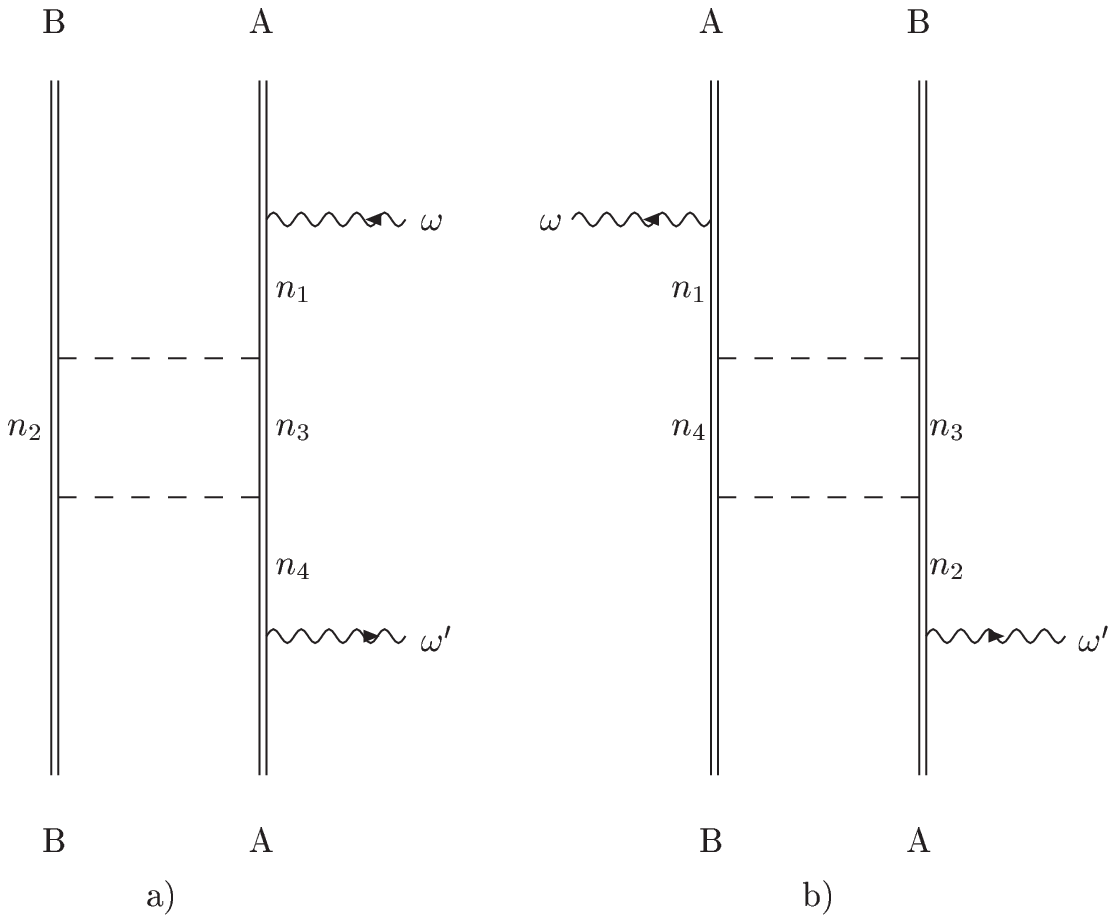}
\caption{The Feynman graphs that describe the exchange by two Coulomb
photons between atomic electrons in the process of  elastic photon
scattering on an atom.} 
\label{fig18}
\end{figure}
\clearpage
\begin{figure}

\includegraphics[bb = 140 430 700 840,scale=1.1]{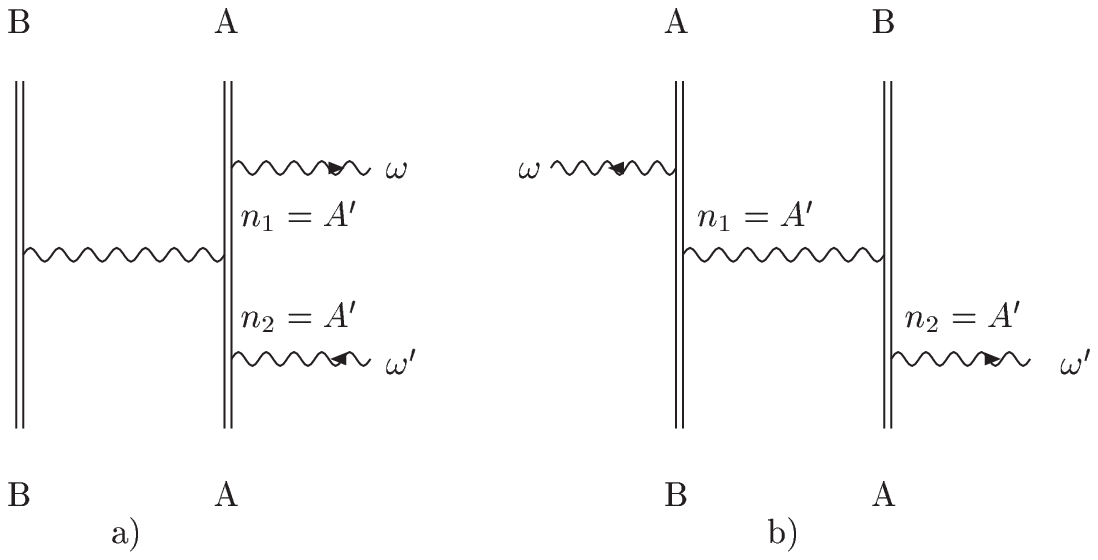}
\caption{ Feynman graphs that describe the elastic photon scattering
on the two-electron ion in  resonance approximation
with the first-order Breit interelectron interaction corrections
taken into account. The wavy internal line denotes the Breit
(transverse) photon. The other notations are the same as in Fig.~17.}
\label{fig19}
\end{figure}
\clearpage
\begin{figure}

\includegraphics[bb = 140 400 550 800,scale=1.1]{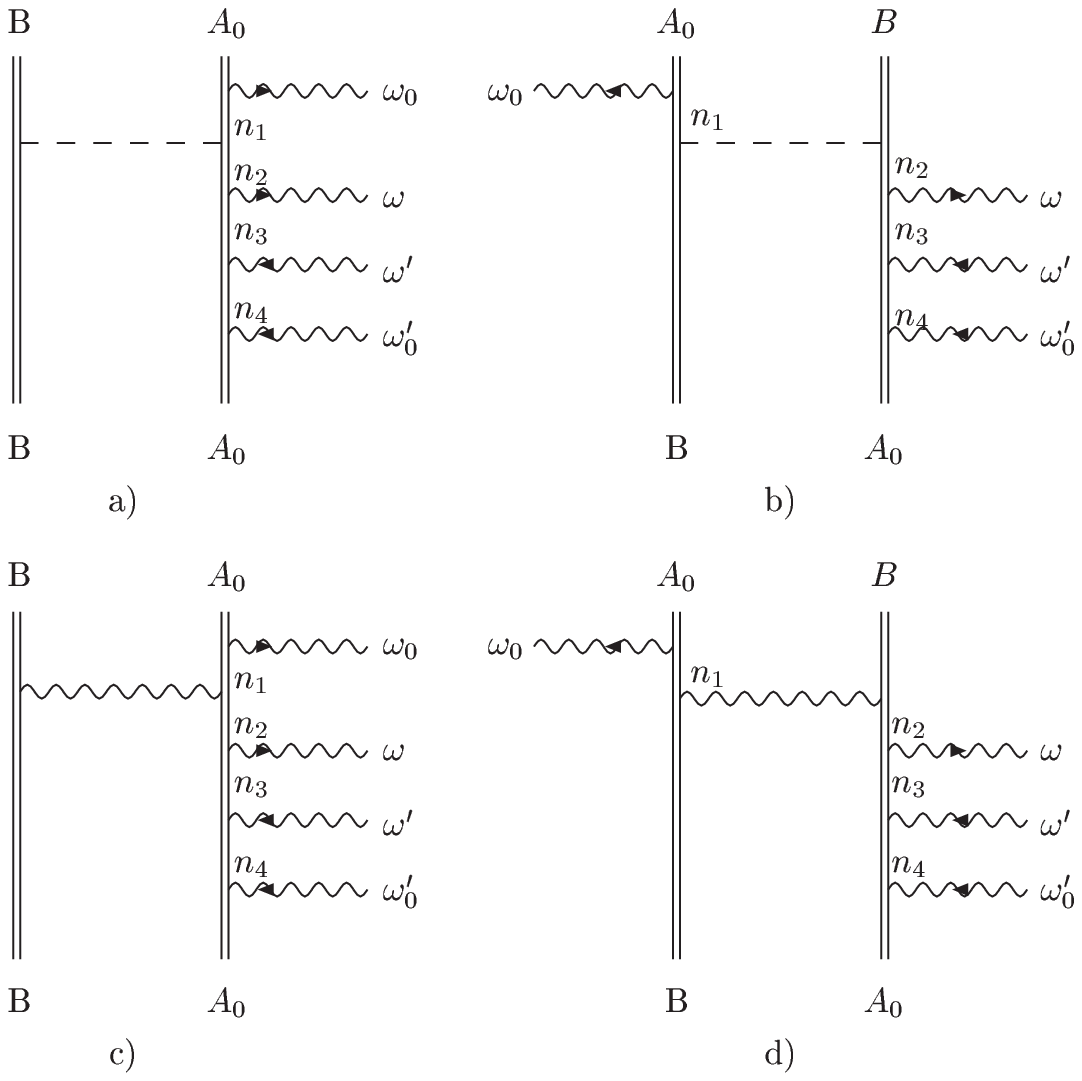} 

\caption{Feynman graphs that describe interelectron interaction first-order
insertions: a), b) correspond to the Coulomb "direct" and
"exchange" graphs; c), d) correspond to the Breit "direct" and
"exchange" graphs.}
\label{fig20} 
\end{figure}
\clearpage
\begin{figure} 

\includegraphics[bb = 140 400 550 800,scale=1.1]{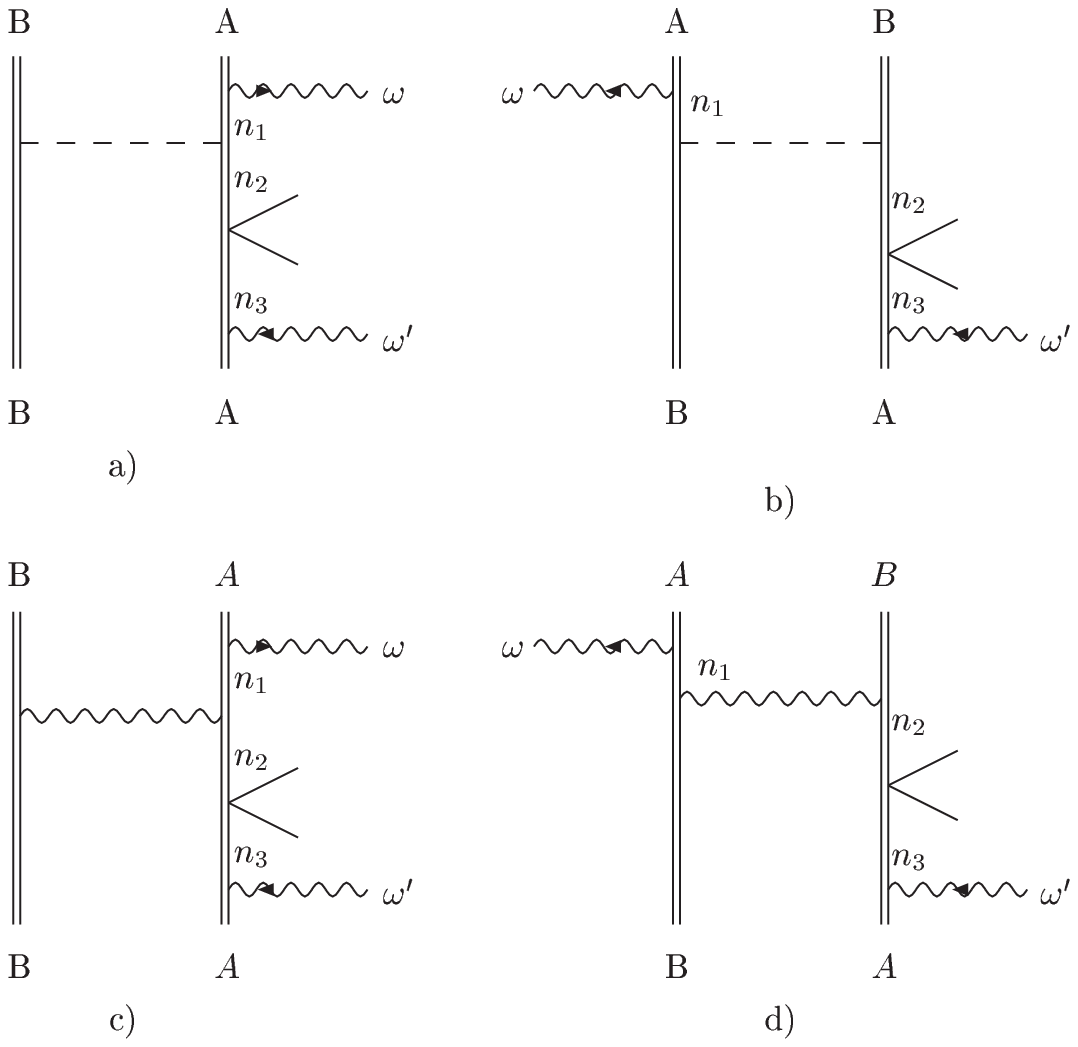}

\caption{Feynman graphs that describe the weak interaction-admixed
amplitude of  photon scattering on two-electron ions. The
graphs a), b) correspond to the interelectron interaction
first-order Coulomb "direct" and "exchange" insertions, the graphs
c), d) correspond to "direct" and "exchange" Breit insertions.} 
\label{fig21}
\end{figure}
\clearpage
\begin{figure}

\includegraphics[bb = 140 400 500 800, scale=1.1]{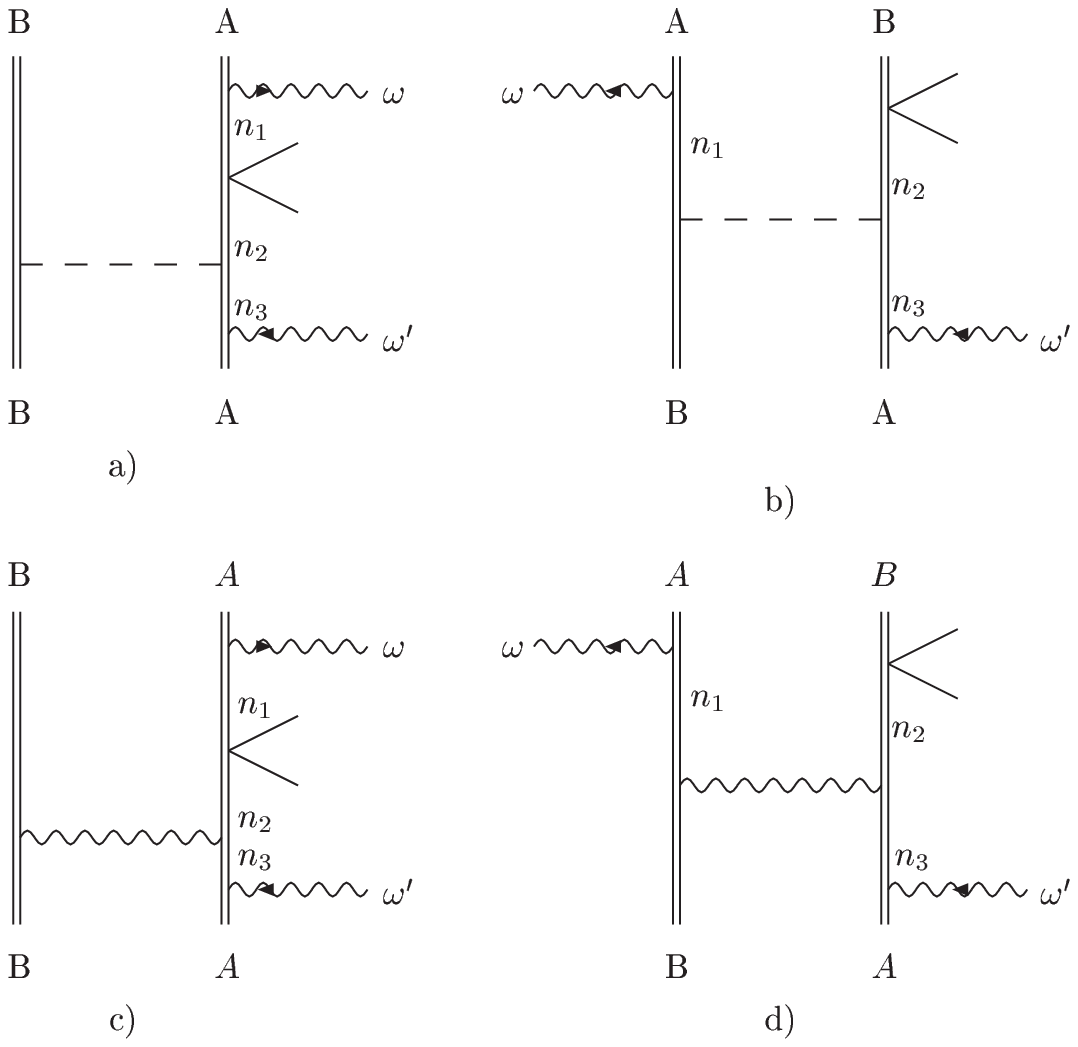}
%{\includegraphics[height=40cm,width=30cm]{figure22.ps}}

\caption{ Feynman graphs that describe the weak interaction-admixed
amplitude of the photon scattering on two-electron ions. Unlike the
graphs in Fig.~21 that correspond to interelectron interaction
corrections insertions to the upper electron propagator in Fig.~10,
Fig.~22 corresponds to the interelectron interaction insertions in the
lower propagator.}
\label{fig22}
\end{figure}
%\clearpage


\begin{thebibliography}{99}
\bibitem{q1}
Th.~St\"{o}hlker, P.~H.~Mokler, F.~Bosch, R.~W.~Dunford,
F.~Franzke, O.~Klepper, C.~Kozhuharov, T.~Ludziejewski,
F.~Nolden, H.~Reich, P.~Rymuza, Z.~Stachura, M.~Steck,
P.~Swiat and A.~Warczak,  {\it Phys. Rev. Lett.}  {\bf 85},  (2000) 3109;\\
P.~H.~Mokler,  {\it Hyperfine Interactions}  {\bf 114},  (1998) 21 ;\\
P.~Seelig, A.~Dax, S.~Faber, M.~Gerlack, G.~Huber,
T.~K\"{u}hl, D.~Marx, P.~Merz, W.~Quint, F.~Schmidt,
H.~Winter and M.~W\"{u}rtz,  {\it Hyperfine Interactions} {\bf 114}, 
(1998) 135;\\ 
P.~Beiersdorfer, A.~L.~Osterfeld, J.~H.~Scofield,
J.~R.~Crespo-Lopez-Urrutia and K.~Widmann,  {\it Phys.~Rev.~Lett.}  {\bf 80},
 (1998) 3022.
\bibitem{qx}
R.~Marrus, A.~Simionovici, P.~Indelicato, D.~D.~Dietrich,
P.~Charles, J.-P.~Briand, K.~Finlayson, F.~Bosh, D.~Liesen and
F.~Parente,  {\it Phys.~ Rev.~Lett.}  {\bf 63}, (1989) 502;\\  
R.~W.~Dunford,
C.~J.~Liu, J.~Last, N.~Berrah-Mansour, R.~Vondrasek,
D.~A.~Church and L.~J.~Curtis,  {\it Phys.~Rev.~A} {\bf 44}, (1991) 764;\\
B.~B.~Birkett, J.~-P.~Briand, P.~Charles, D.~D.~Dietrich,
K.~Finlayson, P.~Indelicato, D.~Liesen, R.~Marrus and
~A.~Simionovic,  {\it Phys.~Rev.~A} {\bf 47}, (1993) R2454;\\ ~A.~Simionovici,
B.~B.~Birkett, J.~-P.~Briand, P.~Charles, D.~D.~Dietrich,
K.~Finlayson, P.~Indelicato, D.~Liesen and R.~Marrus, {\it Phys.~Rev.~A }
{\bf 48}, (1993) 1695.
\bibitem{qx+1}
L.~N.~Labzowsky, G.~L.~Klimchitskaya and Yu.~Yu.~Dmitriev,
       {\it in} ''Relativistic Effects in the Spectra of Atomic System'', IOP Publishing, Bristol, 1993.
\bibitem{qx+2}
P.~J.~Mohr, G.~Plunien and G.~Soff,  {\it Phys. Rep.}  {\bf 293}, (1998) 227.
\bibitem{qx+3}
T.~Beier, {\it Phys. Rep.} {\bf 339}, (2000) 79.
\bibitem{qx+4}
V.~Shabaev, {\it Phys. Rep.} {\bf 356}, (2002) 119.
\bibitem{qx+5}
R.~Marrus and P.~Mohr, {\it Adv. in At. and Mol. Phys.}
 {\bf 14}, (1978) 181.
\bibitem{qy}
P.~Beiersdorfer, A.~L.~Osterfeld and S.~R.~Elliott,  {\it Phys.~Rev.~A}
 {\bf 58}, (1988) 1944.
\bibitem{qy+1}
W.~R.~Johnson and C.~D.~Lin,  {\it Phys.~Rev.}  {\bf 14}, (1976) 565.
\bibitem{qz}
G.~W.~F.~Drake, {\it Phys.~Rev.~A} {\bf 19}, (1979) 1387.
\bibitem{qz+1}
W.~R.~Johnson, D.~Plante and ~J.~Sapirstein,  {\it Adv.~in~ At.~Mol.~Phys.}
 {\bf 46},  (1995) 556.
\bibitem{qz+2}
V.~M.~Shabaev, {\it J.~Phys.~A} {\bf 24}, (1991) 5665. 
\bibitem{qz+3}
V.~Weisskopf and E.~Wigner, {\it Z.~Phys.}  {\bf 63}, (1930) 54.
\bibitem{qz+4}
F.~Low, {\it Phys. Rev.}  {\bf 88},  (1951) 53.
\bibitem{qz+5}
L.~N.~Labzowsky, {\it Zh.~Eksp.~Teor.~Fiz.}  {\bf 85}, (1983) 869.
\bibitem{qz+6}
L.~N.~Labzowsky, {\it J.~Phys.~B}  {\bf 26},  (1993) 1039.
\bibitem{qz+7}
V.~G.~Gorshkov, L.~N.~Labzowsky and A.~A.~Sultanaev, {\it Zh. Eksp. Teor.
Fiz.} {\bf 96}, (1989) 53 [{\it Engl.~Transl.~Sov.~Phys. JETP}  {\bf 69},  (1989) 28].
\bibitem{qz+8}
V.~V.~Karasiev, L.~N.~Labzowsky, A.~V.~Nefiodov, V.~G.~Gorshkov and A.~A.~Sultanaev, {\it Phys.~Scr.~T } {\bf 46}, (1992) 225
\bibitem{qz+9}
L.~N.~Labzowsky, ~V.~V.~Karasiev and ~I.~Goidenko, {\it J.~Phys~B} {\bf 27},
(1994) L439.
\bibitem{qz+10}
L.N.~Labzowsky, I.~Goidenko and D.~Liesen, {\it Phys.~Scr.}  {\bf 56},
(1997) 271.
\bibitem{qz+11}
L.~N.~Labzowsky, D.~A.~Solovyov, G.~Plunien and G.~Soff, {\it 
Phys.~Rev.~Lett.} 
 {\bf 87}, (2001) 143003.
\bibitem{qz+12}
L.~N.~Labzowsky, V.~Karasiev, I.~Lindgren, H.~Persson and S. Salomonson,
{\it Phys.~Scr.~T}  {\bf 46}, (1993) 150
\bibitem{qz+13}
L.~N.~Labzowsky, M.~A.~Tokman, {\it Adv. Quant. Chem.}  {\bf 30},  (1998) 393.
\bibitem{qz+14}
O.~Yu.~Andreev, L.~N.~Labzowsky, G.~Plunien and G.~Soff,
{\it  Phys.~Rev.~A} {\bf 64}, (2001) 042513
\bibitem{qz+15}
V.~G.~Gorshkov and L.~N.~Labzowsky, {\it Pis'ma Zh. Eksp.Teor. Fiz.
} {\bf 19}, (1974) 768 [{\it Engl.~Transl. JETP Lett.}  {\bf 19},  (1974) 394].
\bibitem{qz+16}
A.~Sch\"{a}fer, G.~Soff, P.~Indelicato, B.~M\"{u}ller and W.~Greiner,
{\it Phys.~Rev.~A}  {\bf 40}, (1989) 7362.
\bibitem{qz+17}
G.~von~Oppen, {\it Z.~Phys.~D} {\bf 21}, (1981) 181.
\bibitem{qz+18}
V.~V.~Karasiev, L.~N.~Labzowsky and A.~V.~Nefiodov, {\it Phys.~Lett.~A}  {\bf 172}, 
(1992) 62.
\bibitem{qz+19}
R.~W.~Dunford, {\it Phys.~Rev.~A}  {\bf 54}, (1996) 3820.
\bibitem{qz+20}
M.~Zolotorev and D.~Budker, {\it Phys.~Rev.~ Lett.}  {\bf 78}, (1997) 4717.
\bibitem{qz+21}
L.~N.~Labzowsky, A.~V.~Nefiodov, G.~Plunien, G.~Soff, R.~Marrus and
~D.~Liesen, {\it Phys.~Rev.~A}  {\bf 63}, (2001) 054105.
\bibitem{qz+22}
I.~Bednyakov, L.~N.~Labzowsky, G.~Plunien, G.~Soff and
~V.~Karasiev, {\it ~Phys.~Rev.~A}  {\bf 61}, (1999) 012103
\bibitem{qz+23}
P.~J.~Mohr, Ann. Phys. (N.Y.) {\bf 88}, (1974) 26.
\bibitem{qz+24}
U.~D.~Jentschura, P.~J.~Mohr and G.~Soff, {\it ~Phys.~Rev.~A}  {\bf 63}, (2001) 042512.
\bibitem{qz+25}
H.~Persson, I.~Lindgren and S.~Salomonson, {\it Phys.~Scr.~T}   {\bf 46}, (1993) 125.
\bibitem{qz+26}
H.~M.~Quiney and I.~P.~Grant, {\it Phys.~Scr.~T}  {\bf 46}, (1993) 132.
\bibitem{qz+27}
L.~Labzowsky, I.~Goidenko and A.~Nefiodov, {\it J.~Phys.~B}  {\bf 31}, (1998) L477.
\bibitem{qz+28}
Yu.~Dmitriev, T.~Fedorova and D.~Bogdanov, {\it Phys.~Lett.~A}  {\bf 241}, 
(1998) 84.
\bibitem{qz+29}
E.~A.~Uehling, {\it Phys.~Rev.}  {\bf 48}, (1935) 55.
\bibitem{qz+30}
L.~N.~Labzowsky, A.~O.~Mitrushenkov, {\it Phys.~Rev.~A}  {\bf 53}, (1996) 3029.
\bibitem{qz+31}
E.~Lifshits, L.~Landau and L.~Pitajevskii, ``Quantum Electrodynamics'', 
Pergamon Press, London-Paris,
1983. 
\bibitem{qz+32}
C.~Caso et al., {\it Eur.~Phys.~J.~C}   {\bf 3},  (1998) 1.
\bibitem{qz+33}
S.~Blundell, P.~J.~Mohr, W.~R.~Johnson and J.~Sapirstein, {\it Phys.~Rev.~A } {\bf 48},
 (1993) 2615.
\bibitem{qz+34}
I.~Lindgren, H.~Persson, S.~Salomonson and L.~Labzowsky, 
{\it Phys.~Rev.~A} {\bf 51}, 
 (1995) 1167.
\bibitem{qz+35}
V.~A.~Yerokhin, A.~N.~Artemyev, V.~M.~Shabaev, M.~M.~Sysak, O.~M.~Zherebtsov
and G.~Soff, {\it Phys.~Rev.~Lett. }  {\bf 85}, (2000) 4699.
\bibitem{qz+36}
V.~A.~Yerokhin, A.~N.~Artemyev, V.~M.~Shabaev, M.~M.~Sysak, O.~M.~Zherebtsov
and ~G.~Soff, {\it Phys.~Rev.~A} {\bf 64}, (2001) 032109
\bibitem{qz+37}
P.~J.~Mohr and J.~Sapirstein, {\it Phys.~Rev.~A}  {\bf 62},  (2000) 052501.
\bibitem{qz+39}
L.~N.~Labzowsky and M.~A.~Tokman, {\it J.~Phys.~B}  {\bf 28}, (1995) 3717.
\bibitem{qz+40}
L.~N.~Labzowsky and M.~A.~Tokman, {\it Adv.~Quant.~Chem.}  {\bf 30},  (1998) 393.
\end{thebibliography}
\end{document}